\documentclass[useAMS,usenatbib]{mn2e}
\usepackage{epsfig}

\title[IC1396W]{A near-infrared variability study in the cloud IC1396W: low star-forming efficiency
and two new eclipsing binaries}

\author[Alexander Scholz et al.]{Alexander Scholz$^{1,5}$\thanks{E-mail:
aleks@cp.dias.ie}, Dirk Froebrich$^{2}$, Chris J. Davis$^{3}$, Helmut Meusinger$^{4}$\\
$^{1}$School of Cosmic Physics, Dublin Institute for Advanced Studies, 31 Fitzwilliam Place, Dublin 2, Ireland\\
$^{2}$Centre for Astrophysics and Planetary Science, University of Kent, Canterbury, CT2 7NH, United Kingdom\\
$^{3}$Joint Astronomy Centre, 660 North A'ohoku Place, University Park, Hilo, Hawaii 96720, United States of America\\
$^{4}$Th{\"u}ringer Landessternwarte Tautenburg, Sternwarte 5, D-07778 Tautenburg, Germany\\
$^{5}$SUPA, School of Physics \& Astronomy, University of St. Andrews, North Haugh, St. Andrews, KY16 9SS, United Kingdom
}

\begin{document}

\date{Accepted. Received.}

\pagerange{\pageref{firstpage}--\pageref{lastpage}} \pubyear{2002}

\maketitle

\label{firstpage}

\begin{abstract}
Identifying the population of young stellar objects (YSOs) in high extinction regions is a
prerequisite for studies of star formation. This task is not trivial, as reddened background
objects can be indistinguishable from YSOs in near-infrared colour-colour diagrams. Here we 
combine deep JHK photometry with J- and K-band lightcurves, obtained with UKIRT/WFCAM, to 
explore the YSO population in the dark cloud IC1396W. We demonstrate that a colour-variability
criterion can provide useful constraints on the star forming activity in embedded regions.
For IC1396W we find that a near-infrared colour analysis alone vastly overestimates the number of YSOs. 
In total, the globule probably harbours not more than ten YSOs, among them a system of two young
stars embedded in a small ($\sim 10000$\,AU) reflection nebula. This translates into a star 
forming efficiency SFE of $\sim 1$\%, which is low compared with nearby more massive star forming regions, 
but similar to less massive globules. We confirm that IC1396W is likely associated with the IC1396 HII region. 
One possible explanation for the low SFE is the relatively large distance to the ionizing O-star in
the central part of IC1396. Serendipitously, our variability campaign yields two new eclipsing 
binaries, and eight periodic variables, most of them with the characteristics of contact binaries.
\end{abstract}


\section{Introduction}

Our knowledge about the origin of low-mass stars is primarily based on observations of a few ($\sim 10$)
nearby star forming regions. Although many of these regions are quite similar (for example, most of them
do not harbour massive stars), there are indications for environmental differences in the
outcome of star formation \citep[e.g.][]{2000prpl.conf..121M,2008ApJ...688..377S,2009ApJ...703..399L}. 
Moreover, most of the extensively studied clouds are similar in star formation efficiency \citep{2009ApJS..181..321E}, 
indicating a comparable evolutionary state \citep[see][for an exceptional case]{2009ApJ...704..292F}
Hence, there is an incentive to extend our observational studies to more distant and diverse 
regions. 

The problem is to find efficient means to identify clusters of embedded T Tauri stars. Since extinction 
is often too high for optical observations, the surveys are best carried out in the near/mid-infrared. 
The established method is to look for excess emission from circumstellar disks using colour-colour diagrams.
At least in the near-infrared, however, reddenened background giants and extragalactic objects can occupy
the same regions in colour-colour space as young stellar objects (YSOs), as extinction and disks can 
redden the colour in a similar way. Also, this method will only probe for classical T Tauri stars with 
disks (CTTS) and miss the population of diskless weak-line T Tauri stars (WTTS). 

Thus, in order to select a more complete and less contaminated sample for follow-up spectroscopy, additional
criteria have to be used to separate background/foreground objects from YSOs. One good option here is
variability: T Tauri stars exhibit characteristic and well-studied photometric variations due to spots,
activity, accretion, and disks \citep[e.g.][]{1994AJ....108.1906H,2001AJ....121.3160C,2002AJ....124.1001C,2008A&A...485..155A}, 
which can easily be probed with relatively small telescopes. We present here a case study for a combined 
colour-variability criterion to identify YSOs in a high extinction region.

\begin{figure}
\includegraphics[width=8.5cm,angle=0]{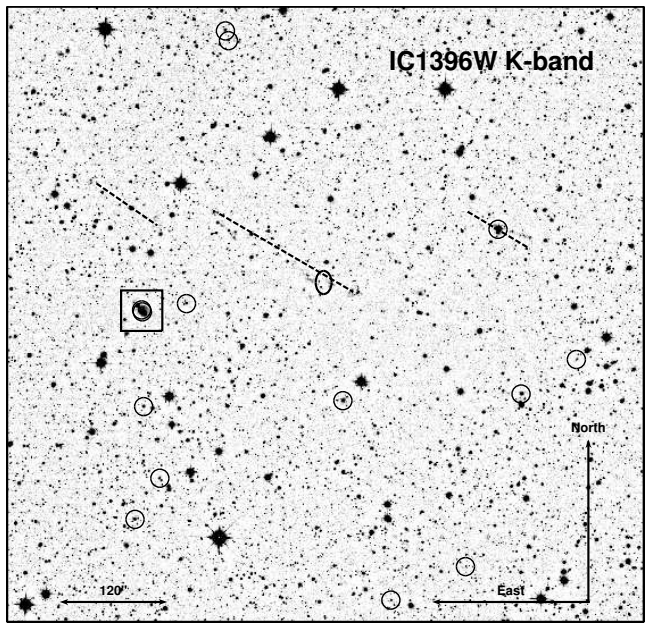} 
\caption{K-band image of the cloud IC1396W. This image has been created by co-adding the individual
K-band frames from our time series observations. The positions and directions of the three outflows are
indicated with dashed lines \citep{2003A&A...407..207F}. The position of IRAS 21246+5743 is marked with
an ellipse. The YSO candidates identified in this paper (see Table \ref{t2}) are marked with circles;
one of them is further north and not seen in the image. 
The square indicates the position of the small nebula with two embedded YSOs, see Sect. \ref{comb} and
Fig. \ref{f10}. 
\label{f100}}
\end{figure}

Our target region is the dark cloud IC1396W\footnote{$\alpha = 21^h26^m$ $\delta = +58^o00^m$, galactic
coordinates $b = 98.3^o$, $l=+05.2^o$}
in the vicinity, but outside, of the large HII region IC1396 
which is ionized by the O6.5V star HD 206267. IC1396 contains several prominent examples of star forming
activity, for example IC1396N \citep{2001A&A...376..271C,2001A&A...376..553N} and IC1396A \citep{2004ApJS..154..385R}, 
see \citet{2005A&A...432..575F} for a summary. If IC1396W is associated with this region, the distance would 
be $\sim 750$\,pc \citep{1979A&A....75..345M}, but so far no independent distance measurement is available. Given its 
relatively simple structure and radiation environment, the sample of globules in IC1396 is an ideal 
testbed to probe the effect of an ionizing star on the surrounding gas and embedded star forming activity. 
The available data is consistent with a scenario in which the radiation from the O star impacts both the 
globule masses and the star forming activity in the globules: With progressively larger distances from the 
ionizing star, the globule masses increase, while the star forming activity drops \citep{2005A&A...432..575F}.

IC1396W contains the red source IRAS 21246+5743 and at least three molecular hydrogen outflows 
\citep[Fig. \ref{f100},][]{2003A&A...407..207F}. Associated with these outflows are the features 
MHO 2267-2772 in the recently 
released Catalogue of Molecular Hydrogen Emission-Line Objects \citep{2010A&A...511A..24D}.
Several of the H$_2$ emission features are detected in the optical as Herbig-Haro 
objects HH 864A-C by \citet{2005A&A...432..575F}. The same authors estimate the cloud 
mass to be about 500$\,M_{\odot}$ and the size to be $\sim 2$\,pc, based on the distance of 750\,pc, which makes 
it one of the largest clouds in the IC1396 region. The IRAS source seems to be a young Class-0 object of 
about 16$\,L_{\odot}$ and 33\,K \citep{2003MNRAS.346..163F} and also the driving source of the main outflow 
HH864. The ISO maps at 160 and 200$\,\mu m$ give evidence for the presence of two more 
embedded sources close to the central object ($2.5'$ SW and NE, respectively). Thus, this is a site 
of active, ongoing star formation.\footnote{CO(1-0) observations by \citet{2006NewA...12..111Z} do not detect the 
outflows nor dense cores associated with the possible driving sources in the cloud. Their beam size of 
$106"\times 70"$, however, is much larger than typical Class-0 sources, i.e. insufficient spatial resolution 
can explain this result.}

In a shallow near-infrared survey, we found that IC1396W harbours a population of reddened objects, most of them 
clustered in a region coinciding with a clump of gas \citep{2003A&A...407..207F}. We present here new 
near-infrared observations to constrain the number and characteristics of young stars in this globule. 
The idea is to combine colours and variability in the near-infrared as indicator for youth.
After discussing our observations and data reduction in Sect. \ref{data}, we search for YSOs in this region
based on colour, variability, and complementary H$\alpha$ narrow band photometry in Sect. \ref{ident}. As
it turns out, IC1396W harbours only a small number ($\la 10$) of YSOs, indicating a low star formation
efficiency (see Sect. \ref{yso}). Our variability database yields a number of serendipitous discoveries,
which are discussed in Sect. \ref{seren}.

\section{Data acquisition and analysis}
\label{data}

\subsection{Observations}

We observed the cloud using the WFCAM instrument (Casali et al. 2007) at the UK Infra-Red
Telescope. WFCAM houses 4 Rockwell Hawaii-II (HgCdTe 2048x2048) arrays spaced by 94\% in the 
focal plane. A fixed optical auto-guider array is installed between these four arrays.  The 
near-IR pixel size is 0\farcs4; consequently a single exposure covers 0.21 square degrees, 
with each chip covering $13.65'\times13.65'$. WFCAM contains 8 filter paddles, one of which 
holds blanks for taking dark frames. We used the J- ($\lambda = 1.25\,\mu m$, $\Delta\lambda=0.16\,\mu m$), 
H- ($\lambda = 1.64\,\mu m$, $\Delta\lambda=0.29\,\mu m$), and K- ($\lambda = 2.20\,\mu m$, $\Delta\lambda=0.34\,\mu m$)
band filters. These filters meet the specifications of the Mauna Kea Consortium filter set 
\citep{2002PASP..114..180T}.

Time series observations alternately in the J- and K-band were obtained in the three nights 3-5 July 2006.
In each of these three nights, the field was monitored over $\sim 5$\,h.
Additionally the field was observed in the H-band on July 2 2006. The seeing throughout these nights
was mostly 0\farcs8-1\farcs1. In particular the first night (July 2) is affected by variable 
transparency due to cirrus. 

The cloud IC1396W is small enough to fit into one WFCAM detector chip (see Fig. \ref{f100}). 
We pointed detector 1 (the SW chip) 
to the target; the other three chips cover adjacent regions without evidence for star formation. 
A 5-point jitter was used in all observations. The detector was stepped in RA and DEC by integer multiples 
of 3\farcs2 or 6\farcs4. These steps represent an integer number of WFCAM IR array and autoguider CCD pixels. 
It also allows us to compensate for bad pixels. Additionally we used a "2x2-small" microstep pattern with 
offsets of 1\farcs4. This results in proper sampling of the seeing with WFCAM's 0\farcs4 pixels. 
 
At each jitter position we integrated for $5 \times 5$\,sec. With the 20 positions (dithering and micro-stepping)
we obtain an per pixel exposure time for each time series image of 500\,sec in J- and K-band. There 
are 39 time series frames in J (13, 12, 14 in the three nights) and 38 in K (13, 11, 14). Additionally 
we obtained 8 images in the H-band with the same per pixel exposure time. Thus, for the co-added 
image of the region we have a per pixel integration time of 5.4\,h in J, 1.1\,h in H and 5.3\,h in K.

To complement the near-infrared dataset, IC1396W was observed in the optical wavelength 
regime using the 2\,m telescope of the Th{\"u}ringer Landessternwarte Tautenburg (TLS) in its 
Schmidt mode. These observations were carried out with the 2k $\times$ 2k SITe CCD of $24 \times 24\,\mu m$
pixel size, corresponding to a $42' \times 42'$ field of view, on December 7th 2007. The 
field was imaged in the R-band, a narrow-band filter centered on the H$\alpha$ emission line 
(6562\,\AA), and another narrow-band filter centered on 6673\,\AA. In the R band,
three exposures were taken with an integration time of 100\,sec each. The narrow band 
images were exposed 300\,sec each. For calibration purposes domeflats and bias frames were 
included as well. 

\subsection{Image reduction and stacking}

All WFCAM data were reduced by the Cambridge Astronomical Survey Unit (CASU) and are distributed through 
a dedicated archive hosted by the Wide Field Astronomy Unit (WFAU) in Edinburgh. The CASU reduction steps 
are described in detail by \citet{2006MNRAS.372.1227D}; these include dark subtraction, bad pixel masking, 
flat fielding using twilight flats, and sky subtraction using a running average of frames obtained before 
and after each exposure. During this process the images are resampled to a scale of 0\farcs2/pixel. 
Astrometric and photometric calibrations are achieved using 2MASS point sources in each image 
\citep{2006MNRAS.372.1227D,2006MNRAS.367..454H}. We downloaded the fully-reduced images from WFAU. 
One reduced Multi-Extension Fits (MEF) file is available for each of the 39 (J), 38 (K) and 8 (H) 
time-series datasets; data from the four arrays in WFCAM are presented as separate images in each
MEF.

To facilitate stacking and time series photometry, all images were co-centred using bright stars. 
The images were also cut so that only positions in the field covered in every individual image in the 
time series (J, H and K) are left. To obtain a deep image in each of the three bands we co-added the 
individual frames. Since the weather conditions were slighly variable, stars have a different 
signal-to-noise ratio in the images. Hence, each co-added frame was given a weight proportional 
to the ratio of flux divided by noise squared. This ensures that the resulting co-added image 
has the maximum signal to noise ratio \citep[e.g.][]{2000A&AS..145..229F}. A coordinate
system was matched to the stacked frame based on 2MASS objects in the field.

For the optical images from the TLS a standard reduction was carried out including 
the subtraction of an average bias frame as well as dividing by the normalized average 
flatfield.

\subsection{Photometry and calibration}
\label{phot}

We produced a raw object catalogue based on the stacked K-band images for detector 1
(SW chip, the IC1396W field) and detector 4 (NW chip, $\sim 13'$N of IC1396W, comparison field). 
This is done with the SExtractor software \citep{1996A&AS..117..393B} with default parameters. 
From this catalogue we rejected all objects with FWHM larger than 8 pixels and elongation 
$A/B>2$. This excludes a number of artefacts, barely resolved binaries, and galaxies. 
Still, the final catalogue does contain a significant fraction of extended objects.

The photometry for all time series images in J- and K-band as well as for the stacked images
in JHK was carried out with the {\tt daophot-phot} routine within IRAF. We kept a constant 
aperture of 10 pixels for all images. significantly larger than the typical seeing of 5 pixels. 
The sky background was estimated from a annulus with an inner radius of 15 pixels and 
5 pixels width. 

The photometry of the stacked frames was shifted to a standard system using $\sim 200$ 
bright stars in the field which are covered by 2MASS. We find zeropoints of 0.97, 1.19, and
0.60\,mag for J-, H- and K-band, respectively. In addition, our photometry shows small
colour terms with coefficients in the range of 0.03\,mag in J- and H-band. We did not attempt 
an absolute calibration for the optical images.

For the time series photometry, a relative calibration is needed to account for the
effects of variable transparency and variable seeing. This is done by deriving a reference
lightcurve from a set of $\sim 150$ bright, non-variable stars for each field and filter, 
selected using the procedure outlined in \citet{2004A&A...419..249S}. In short, from a large 
sample of several hundred lightcurves we selected the ones which show the minimum deviations 
from the ensemble lightcurve. These are averaged to define the reference lightcurve, which 
reflects the variations due to atmospheric effects. This reference lightcurve is subtracted
from all lightcurves to correct for these effects.

\section{Identifying young stellar objects}
\label{ident}

In the following section we will develop a strategy to find young stellar objects based
on the combined information from near-infrared colours and variability. A few benefits and
caveats of this method should be pointed out. 

A colour criterion essentially probes for the
presence of a hot, inner disk, which causes colour excess particularly in $H-K$. A selection
based on near-infrared colour is expected to be strongly contaminated. As discussed for example 
in \citet{2006ApJS..167..256R}, YSOs are expected to populate the area close or below the 
reddening band, and thus overlap in their colours with reddened background objects. A 
near-infrared colour selection is also incomplete as it misses objects without disks (weak-line
T Tauri stars WTTS) or with inner disk holes.

\begin{figure*}
\includegraphics[width=6cm,angle=-90]{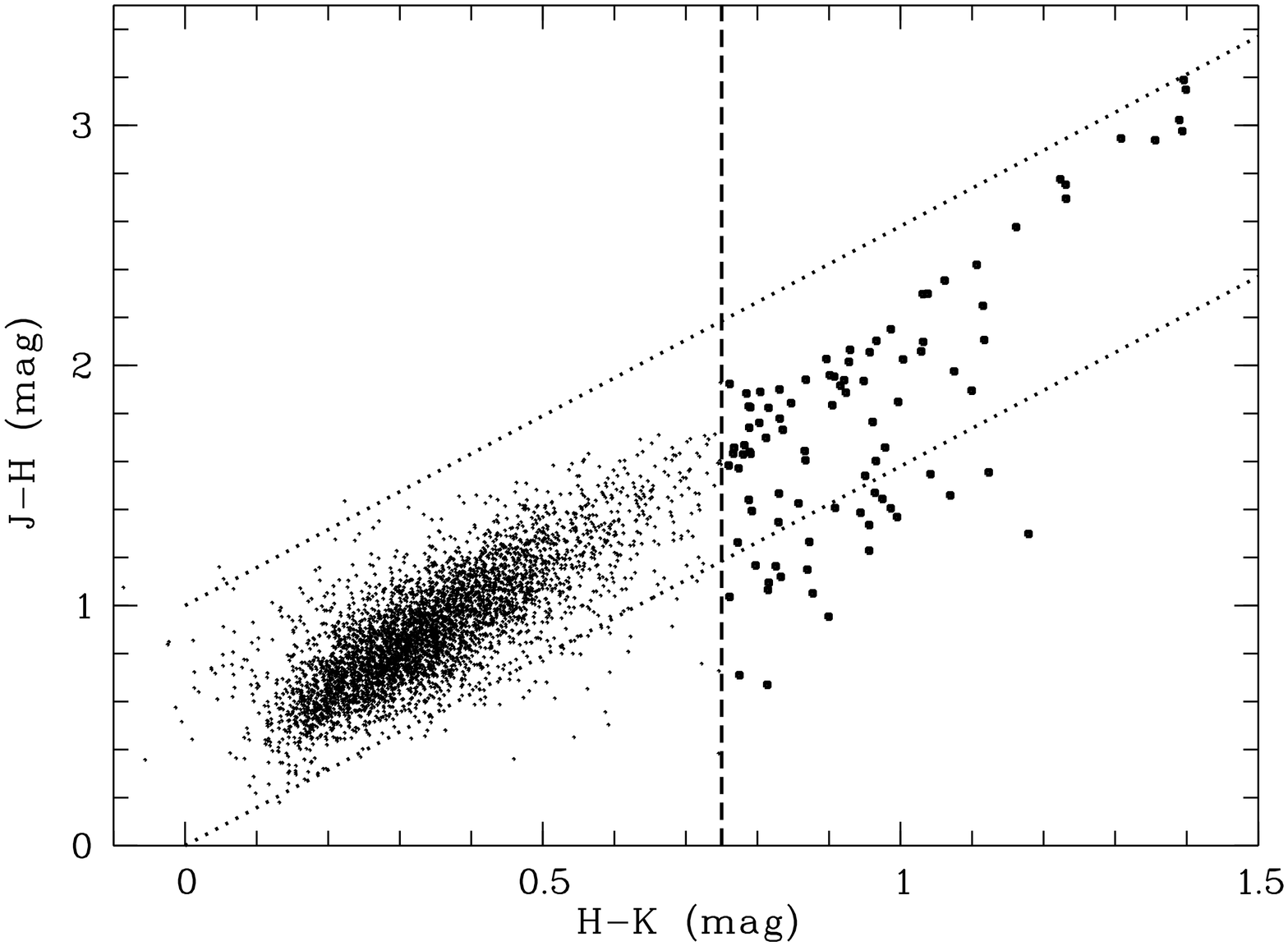} 
\includegraphics[width=6cm,angle=-90]{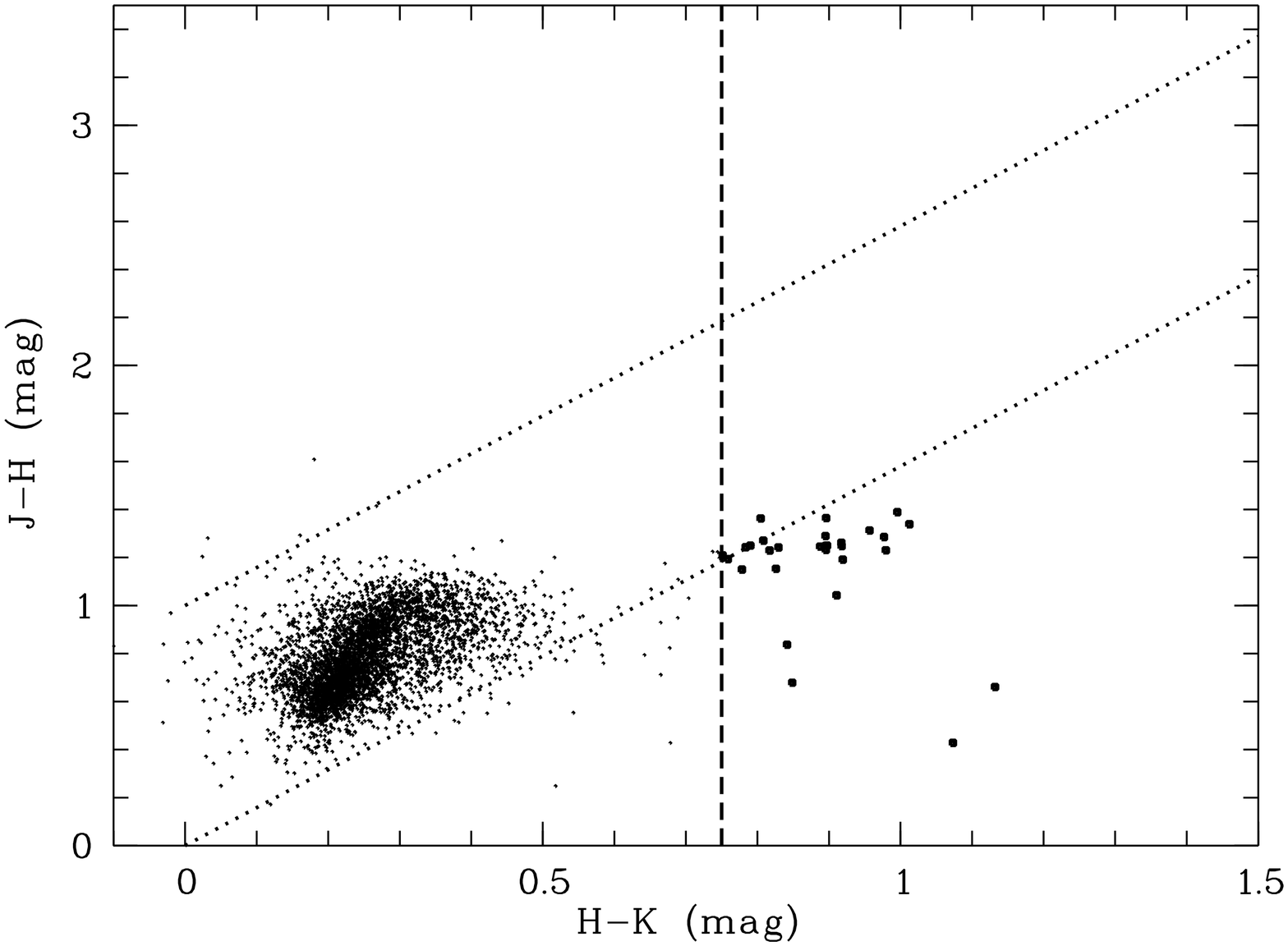} 
\caption{Colour-colour-diagram for the globule IC1396W (left panel) and a comparison
field (right panel). For the colour criterium, we require the objects to have 
$H-K>0.75$. One datapoint at $H-K=1.9$ and $J-H=4.0$ is not plotted.
The IC1396W field has 97 objects fulfilling the criterium. In the comparison field, 
29 objects fulfill this limit, most of them are not point sources. The dotted lines
show the reddening path over $A_V \sim 20$\,mag \citep{1990ARA&A..28...37M}.
\label{f1}}
\end{figure*}

Variability in YSOs can be caused broadly by the presence of an accretion disk (hot spots or variable 
emission/absorption from the disk) and by magnetic activity (cool spots or flares). These
processes will cause characteristic signatures in the near-infrared lightcurves 
\citep{2008A&A...485..155A,2009MNRAS.398..873S}. As for the colour criterion, a YSO selection 
exclusively based on variability will be contaminated and incomplete. Based on the lightcurves it 
is not always possible to separate YSOs from field stars (e.g., magnetically active M dwarfs or 
short-period pulsating giants). This selection will miss YSOs without measurable variability, 
for example the ones without accretion and no detectable spot activity.

A combined colour-variability criterion can reliably establish a sample of YSOs with
little contamination, but it will still not provide a complete census. Specifically it 
will miss 1) deeply embedded sources not visible in the near-infrared bands, 2) objects with 
small inner disk holes and thus no near-infrared excess, 3) WTTS without disk, 4) YSOs without 
measurable variability on the timescales of the observations. The first two 
groups can be identified with further photometric observations at longer wavelengths. WTTS 
can often be found based on variability alone, as they show periodic lightcurves on 
timescales of days and/or flare activity. 

Group 4 may be of major relevance for our study, as our observations only cover three days: 
As discussed in \citet{2009MNRAS.400..603P}, such relatively short runs might miss more than
half of the variables in star forming regions. On timescales of several years, however, almost
100\% of YSOs are found to be variable. This applies to variability amplitudes in the range
of a few percent or larger, which is comparable to our detection limit. By combining our observations 
with 2MASS data we will look for missing objects with long-term variations (Sect. \ref{2mass}).

On the other hand, the fact that our run covers only a few nights helps us to rule out some 
classes of objects that are common contaminants in YSO surveys. For example, quasars may appear 
red, but their dominant variability modes have characteristic timescales in the order of years 
in the quasar rest frame, thus is unlikely to be found with the current dataset. The same 
argument applies to long-periodic pulsating red giants (e.g. Mira variables).

In summary, using colour or variability as observational indicators for youth gives incomplete
and contaminated samples. Combining the two can potentially provide a much cleaner 
sample of YSO candidates, but will not completely solve the incompleteness issue.

\subsection{The colour-colour diagram}
\label{ccd}

From the calibrated photometry of the stacked images (see Sect. \ref{data}) a deep (H-K, J-H) 
colour-colour diagram was constructed, as commonly used to identify objects with disks. This
was done for the field containing IC1396W in chip 1 as well as for the field covered by chip 4
for comparison. The full source catalogue for the IC1396W field has 8211 entries from which 
7965 have reliable photometry in all three bands. Clearly the majority of these objects will
not be young members of IC1396W.\footnote{In the following we use the running number in these 
catalogues to identify individual objects.}

A further cut was made by limiting the diagram to objects for which useful variability information
is available. We include only sources for which the mean photometric noise, as determined from the 
lightcurves, is $<0.2$\,mag in J- and K-band (see Sect. \ref{var}). This reduces the number of objects to
4397. The second cut imposes essentially a magnitude limit; while the full catalogue extends 
down to $J=22$ and $K=19.5$\,mag, these values change to $J=19.4$\,mag and $K=18.0$\,mag after
the variability information is taken into account. The comparison field has in total 7496 objects
with photometry in all bands. After applying the variability cut a catalogue of 4422 objects is 
obtained, with limiting magnitudes of $J=19.3$ and $K=18.0$\,mag.

The resulting colour-colour diagrams are shown in Fig. \ref{f1}. The comparison field shows a 
clump of objects around (0.2,0.8) with a faint tail of objects with (small) $H-K$ excess. Most
of the objects in this tail appear extended and may be red galaxies. There is no indication of 
strong extinction in the comparison field. The field around IC1396W shows the same clump of objects, 
but here it is stretched out along a well-defined path towards red colours. This is best explained 
by strong and variable extinction in the globule, which affects the colours from all background 
sources. 

The dotted lines in Fig. \ref{f1} shows the standard interstellar reddening law with power law index 
$\beta = 1.7$ \citep{1990ARA&A..28...37M}, which matches well with the vector of the reddening seen 
in this field. This provides reassurance in our colour calibration and shows that IC1396W does not
exhibit an unusual $\beta$, as already shown by \citet{2006MNRAS.369.1901F}. 

\begin{figure*}
\includegraphics[width=6cm,angle=-90]{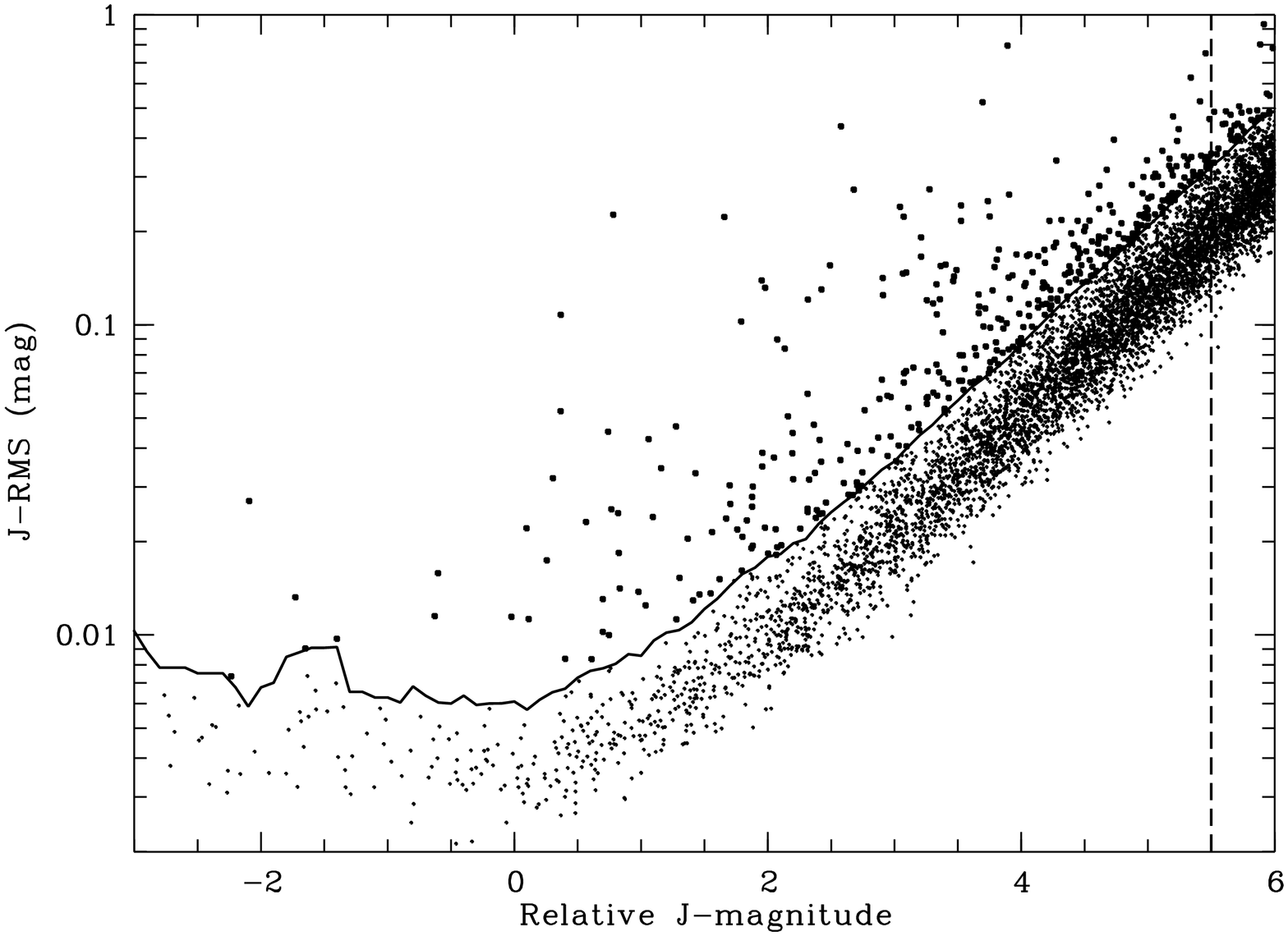} 
\includegraphics[width=6cm,angle=-90]{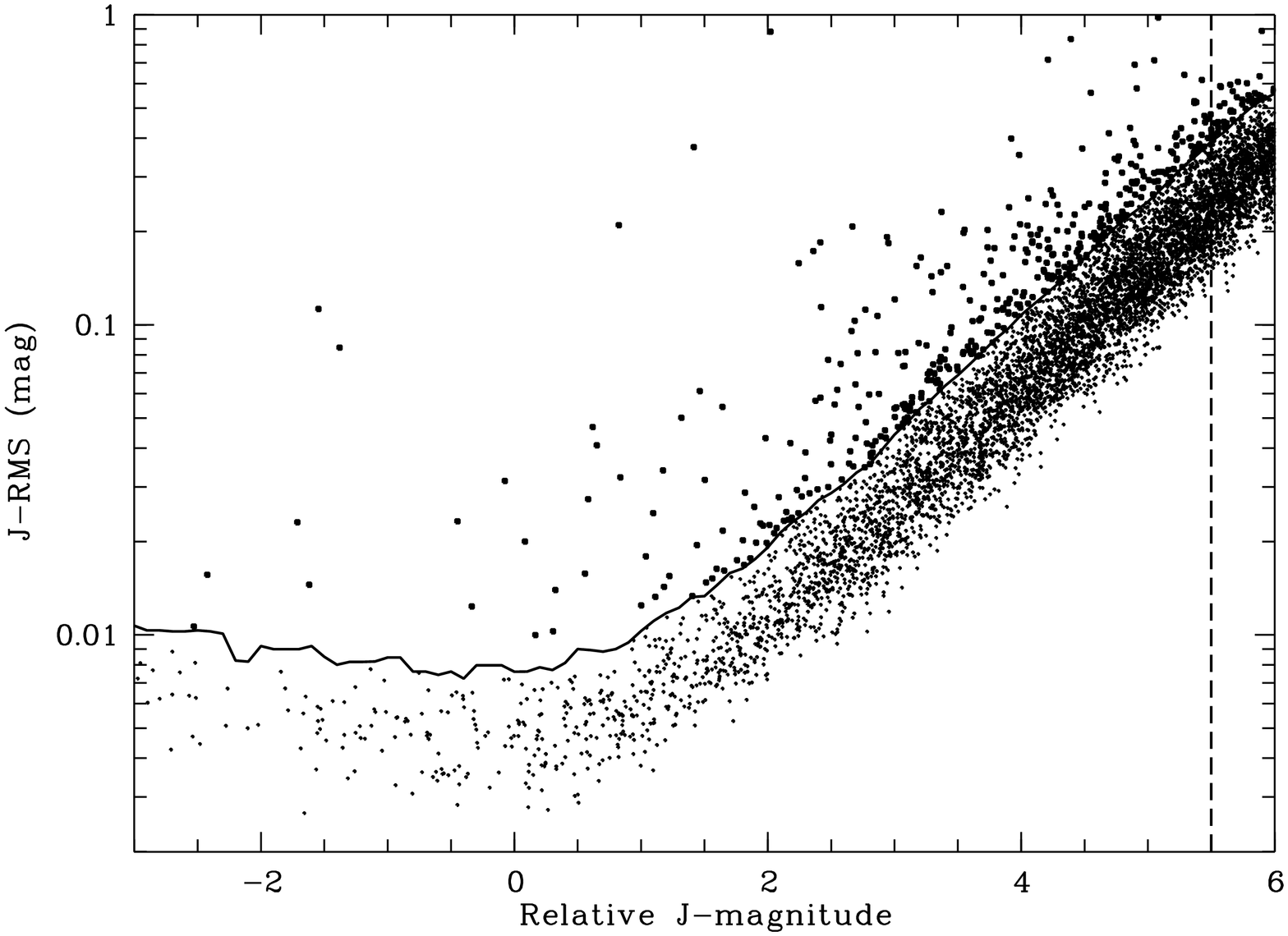}\\ 
\includegraphics[width=6cm,angle=-90]{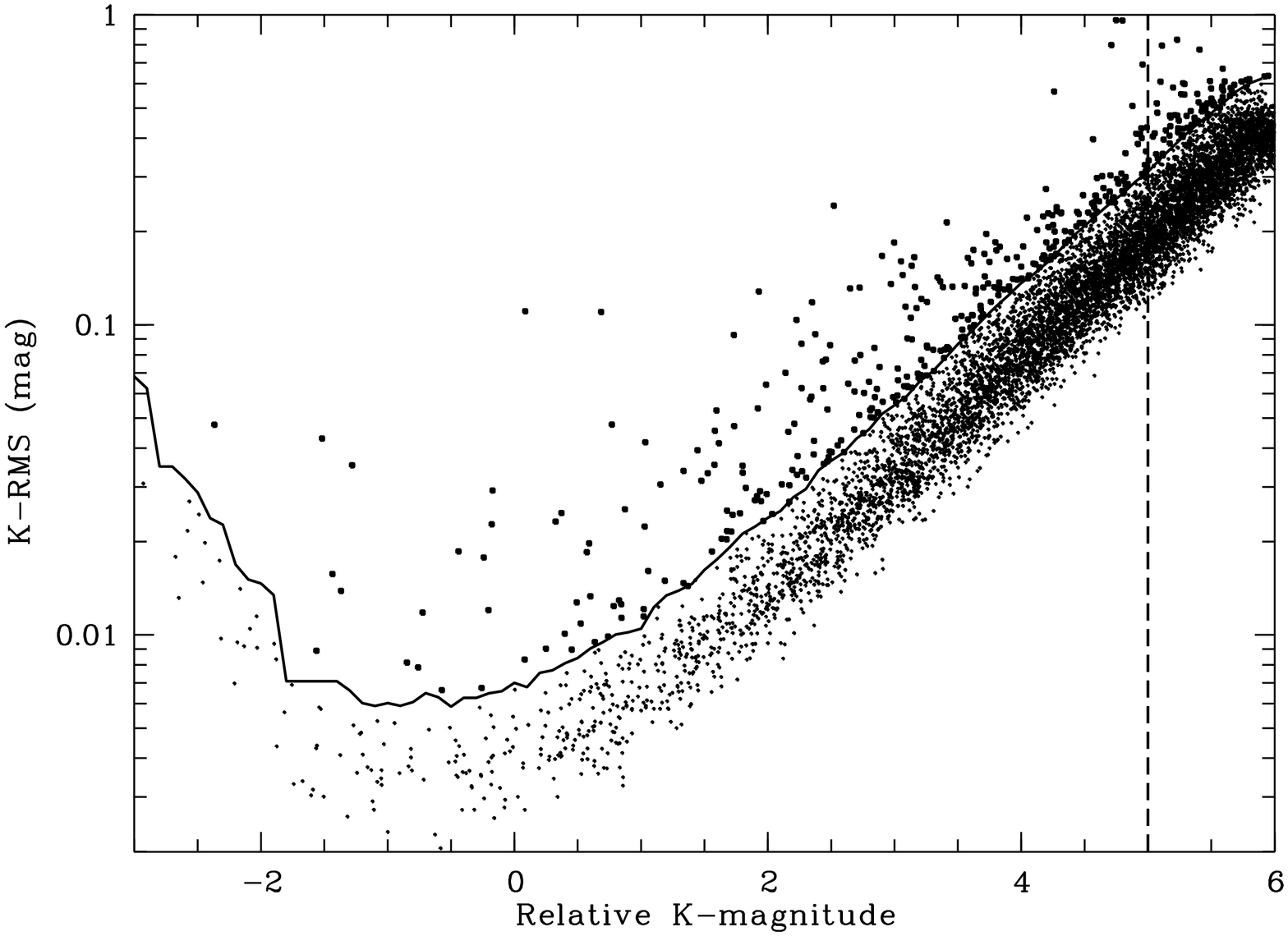} 
\includegraphics[width=6cm,angle=-90]{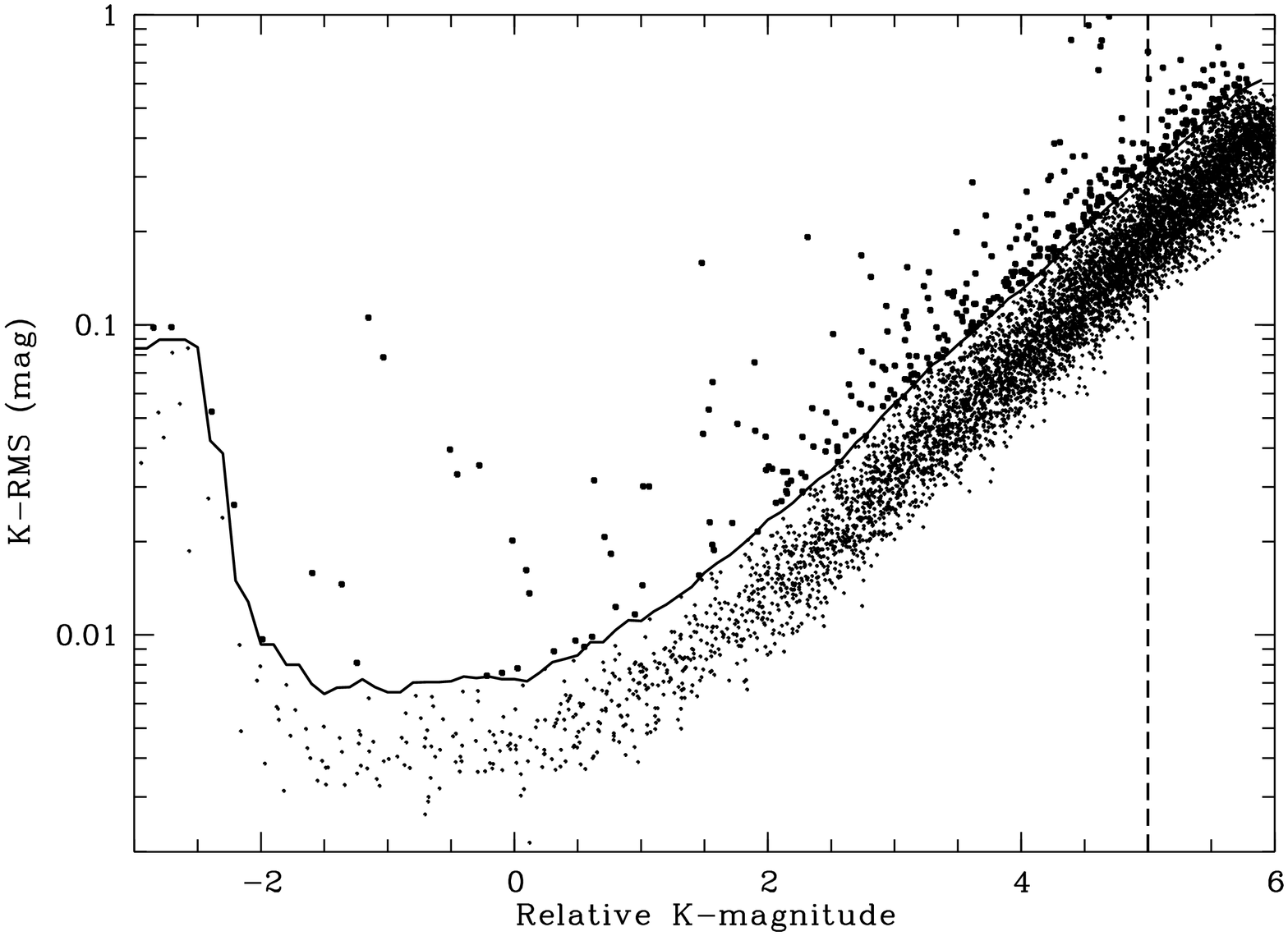} 
\caption{Variability diagram for the globule (left) and for the comparison field (right) in J (upper row) 
and K (lower row). All objects with RMS higher than the solid line are statistically variable ($F>2.6$) 
in comparison with the median RMS of stars with the same magnitude. The threshold is chosen to
minimize the contamination. The zeropoint in relative magnitudes corresponds to $\sim 13.8$  in J-band
and $\sim 12.8$\,mag in K-band. Dashed lines mark the limit where the median RMS exceeds 0.2\,mag \label{f2}}
\end{figure*}

We use these diagrams to select a sample of objects with near-infrared colour excess, which could
be considered candidates for classical T Tauri stars (CTTS). In the following these colour-selected
objects will be called C-sample. CTTS usually occupy a region around or below the reddening vector, 
because they are affected by extinction and additional excess due to circumstellar emission 
\citep{1997AJ....114..288M,2006ApJS..167..256R}. We define a vertical colour cut (dashed line in 
Fig. \ref{f1}) aiming to avoid the bulk of the population in the IC1396W area. The most reasonable 
results are achieved for $H-K>0.75$, which yields a C-sample of 97 objects for IC1396W and 29 for 
the comparison field. Shifting this boundary to the left by 0.1 almost doubles the number of 
objects in the IC1396W area, due to excessive contamination. The C-sample will be contaminated 
by the highly reddenened field stars, but also covers the typical colour range for CTTS. 

\subsection{Variability}
\label{var}
In this section we select variable objects from our UKIRT dataset covering three observing nights.
An object is considered variable if the photometric variations are significantly larger than the
noise expected at this magnitude. This is tested by comparing the root mean square (RMS) with
the median RMS for a given magnitude using a statistical test. The following steps are carried out for
the J- and K-band data separately, in the IC1396W field and in the comparison field. The analysis
starts with the database of lightcurves for all objects in the full source catalogue, 
i.e. 8211 for the IC1396W field.

In a first step we calculate the root mean square (RMS) from the lightcurves of all stars in the 
database. Three iterations of 3$\sigma$ clipping are carried out to exclude outliers. To quantify 
the photometric noise as a function of magnitude, we calculate a running median RMS for 0.5\,mag wide 
bins and 0.1\,mag stepsize from all lightcurves with at least 30 datapoints. This is based on the assumption 
that the overwhelming majority of the stars in our fields are non-variable for our photometric accuracy. 
Strictly speaking, we define 'variable' relative to the bulk of field stars.

This yields an empirical function (see solid lines in Fig. \ref{f2}), it reaches 5\,mmag for bright, unsaturated 
stars and rises exponentially for fainter objects. The increase at the bright end can be attributed to the onset 
of non-linearity. A median RMS of 0.2\,mag is reached at relative magnitudes of 5.5 in the J-band and 5.0 in K-band; 
objects with larger photometric noise are not considered. This cut-off is also used in the analysis of the 
colour-colour diagram (Sect. \ref{ccd}).

For any given star we then compare the actual RMS with the median RMS in the appropriate magnitude
bin. We use the statistical F-test, which tests the zero hypothesis that the variances in two
samples are consistent (where the variance is the square of the RMS). The test quantity $F$ is simply 
the ratio of measured variance vs. median variance; we are looking for lightcurves for which this
quantity exceeds 1.0 significantly. For our lightcurves and assuming that the noise is Gaussian, 
$F \sim 2.2$ corresponds roughly to 1\% likelihood that the zero hypothesis is valid, which is the
false alarm probability for an object to be wrongly identified as variable at this F-level.

We choose a slightly higher threshold of $F=2.6$ to identify the sample of variables, aiming to reduce 
the contamination and to avoid the bulk of the datapoints. In Fig. \ref{f2} we show 
the RMS vs. magnitude diagram for J- and K-band. Overplotted in solid lines is the threshold
for $F=2.6$. All objects above this threshold are marked. For the J-band this gives a sample of 389 
out of 4800 objects with useful lightcurve. For K-band the  variable sample has 320 out of 4878 objects. 
165 objects are variable in both bands (called V-sample in the following).

The same procedure for the comparison field yields 417 variables in J-band, 286 in K-band, and 154
in both bands. The comparable number of variables in the two fields already indicates that IC1396W 
cannot harbour a 
large population of young, variable objects. 

\subsection{Combining colour and variability}
\label{comb}

\begin{figure*}
\includegraphics[width=6cm,angle=-90]{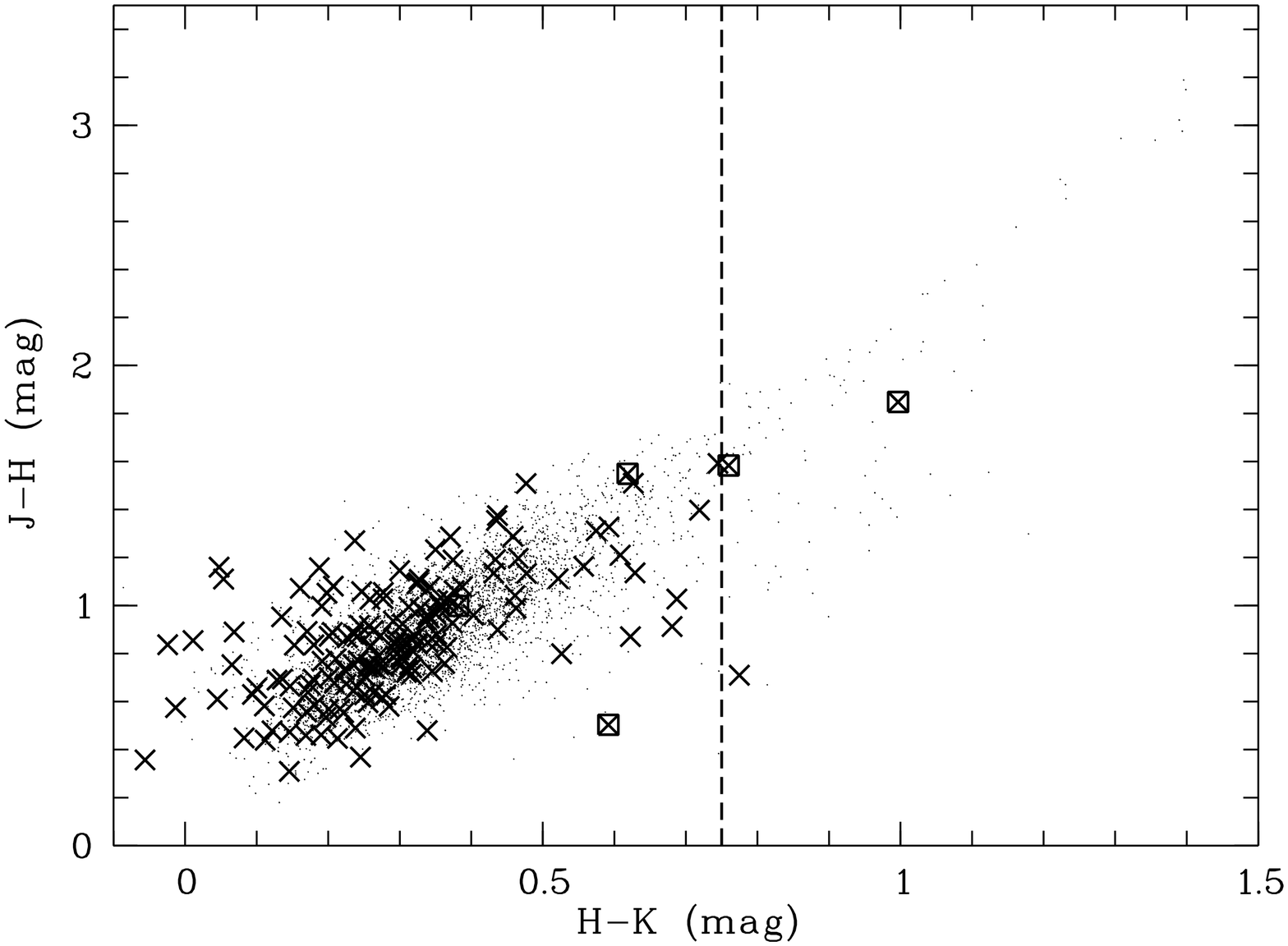} 
\includegraphics[width=6cm,angle=-90]{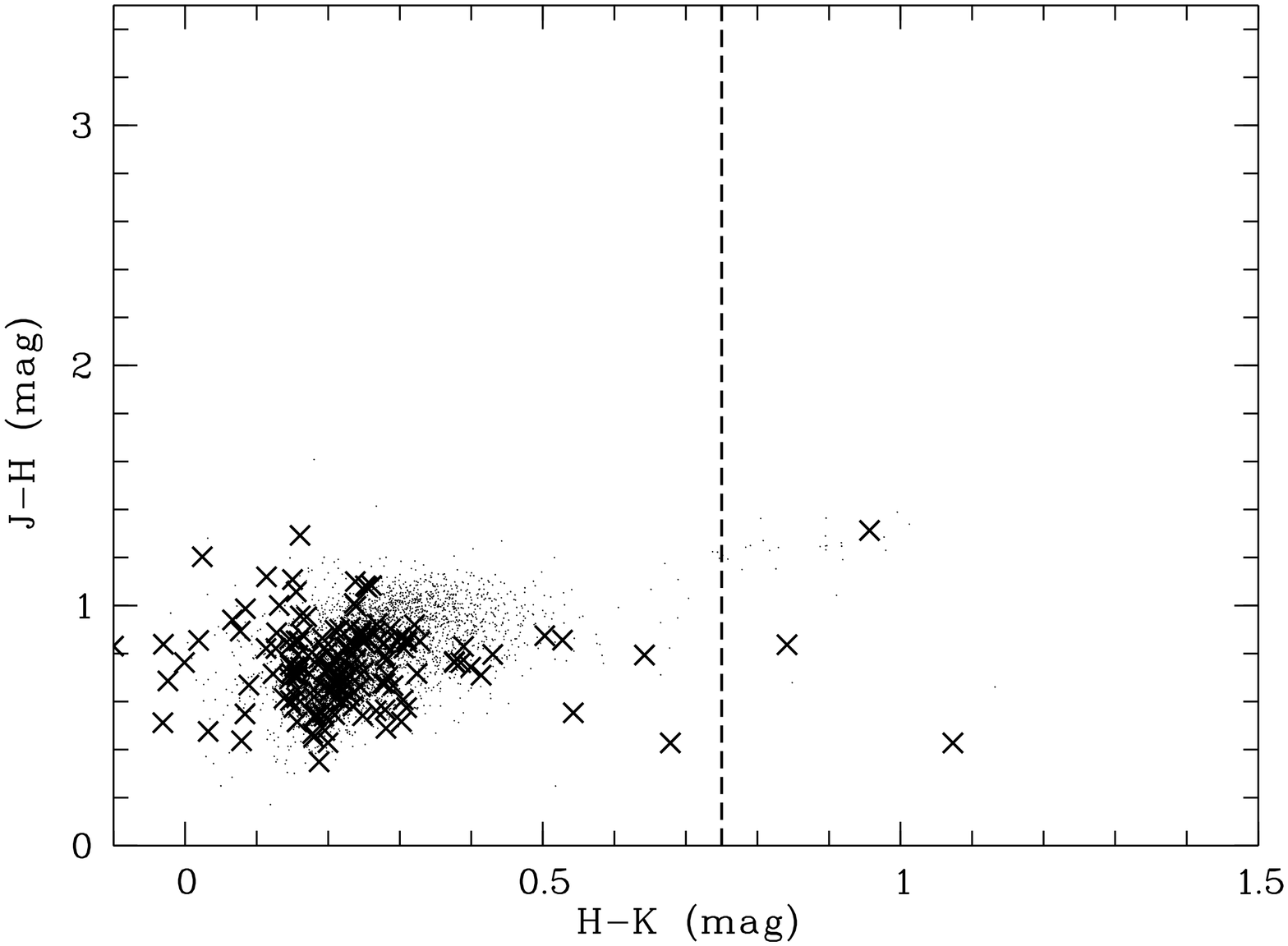} 
\caption{Colour-colour diagram for the globule (left panel) and the comparison field (right panel), variable
objects are marked (crosses: variable in our time series, see Fig. \ref{f2}; squares: variable in comparison
with 2MASS data). The globule has 165 objects variable in K and J. In the comparison field, the same procedure 
gives 154 variable objects. \label{f3}}
\end{figure*}

In Fig. \ref{f3} we combine the information from the two previous subsections and look
for overlap between C- and V-sample. The figures show the colour-colour diagrams for the 
IC1396W and the comparison field, as in Fig. \ref{f1}. All objects from the V-sample
are marked. The adopted threshold for the C-sample is shown as a dashed line. 

Quite obviously there is no significant population of objects in the globule which are
reddened and variable. Most of the sources in the V-sample are scattered over the main 
bulk of datapoints at normal main-sequence colours. The number of red and variable objects
is three in both fields. One of these sources in IC1396W is 12' 
away from the globule core and thus likely contamination. The small overlap between 
red and variable objects argues for only a small population of YSOs in IC1396W.

Two of the sources in the C- and V-sample, 1-2189 and 1-2314, are located close together (2\farcs3), 
are unresolved in previous observations, and seen within a bright nebulosity, which has an entry 
in the 2MASS Extended Source Catalogue\footnote{2MASS ID: 2MASX J21263151+5755526}. It is located in 
a region of strong extinction, evidenced by low object density and red colours. This region has already 
been identified in \citet{2003A&A...407..207F} as SE extension of the IC1396W cloud (see Fig. \ref{f100}). 
All stars in the 
immediate neighbourhood of the nebula are significantly redder than our two candidates. The nebulosity 
is a fuzzy blob and does not show any resolved structure (apart from the point sources). 

\begin{figure}
\center
\includegraphics[width=8cm]{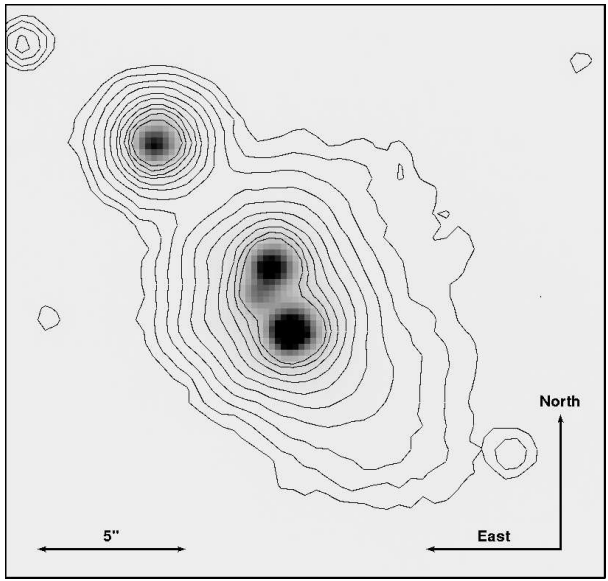} 
\caption{K-band image of two red and variable objects in a nebula, N is up and E is left. The objects 
1-2314 and 1-2189 are seen in the center, together with a faint third source. The object 1-1996 is seen 
towards the NE. Contours are on a log-scale from 10 to 10000 counts. 
\label{f10}}
\end{figure}

As shown in Fig. \ref{f10}, the system actually harbours three point sources: The two red and variable objects, 
both with K-magnitudes of 11.3-11.5\,mag and 2\farcs3 separated in NE-SW direction, plus a faint third object
1\farcs1 away from 1-2314, which is undetected in our catalogue. Another star (1-1996 with $K=12.6$)
is seen to the NE just outside the nebula; with $H-K = 1.4$ and $J-H = 3.0$ this object is among the
reddest in our catalogue with $A_V \sim 15-20$. It might be another embedded YSO or a background star.

The fact that 1-2189 and 1-2314 pass the colour-variability test combined with the
characteristic of the nebula makes it likely that we are dealing with a system of young stellar objects
embedded in a reflection nebula. Specifically, based on the location in a high-extinction area we argue that
this is almost certainly not an extragalactic background object. If the distance of 750\,pc for IC1396W 
is correct (see Sect. \ref{sfe}), the separation for the two stars in the nebulosity is 1700\,AU in 
projection, the diameter of the nebulosity $\sim 10000$\,AU. Thus, this looks like a typical localized 
cell of star formation as they are commonly found for example in the Taurus-Auriga star forming region.

\subsection{Probing long-term variability}
\label{2mass}

As mentioned earlier, the concern is that our relatively short monitoring run does not
allow us to identify objects that are variable on long timescales. The 2MASS data is used
to check for long-term variability for the brighter objects in our catalogue. We find 595
counterparts with photometry in the 2MASS database, taken several years earlier than our
own data. This implies that the test can only be carried out for 14\% of the total sample
considered in the previous sections. In addition, our information about long-term 
variability is limited because 2MASS provides only one epoch. Hence, we might miss 
long-term variables, particularly at the faint end of our survey.

For the overwhelming majority of the objects with 2MASS counterpart the difference between our
magnitudes and the 2MASS values is within the expected photometric errors. For example, the 
deviations in the J-band are $<0.1$\,mag down to $J=15$\,mag.
Five objects were found with clear discrepancy between 2MASS and our magnitude in J- and
K-band. However, all five are already contained in our V-sample for J- as well as K-band, i.e. 
this test does not yield any new candidates and confirms the scarcity of YSOs in IC1396W. In 
Fig. \ref{f3} these five objects are shown with squares. This includes the red sources identified
in the nebulosity (see Sect. \ref{comb}); in these cases however the test
is unreliable, as they are not resolved in 2MASS. The total near-infrared brightness in the
Extended Source Catalogue is comparable to the combined flux from the two point sources.

\subsection{Variability characteristics}
\label{char}

On the timescales covered by our time series, young stellar objects frequently show two types 
of variability: a) strictly periodic with amplitudes $\la 0.2$\,mag, b) partly
irregular variations with large amplitudes often exceeding 0.5\,mag. These two categories 
represent the type I and type II variability identified by \citet{1994AJ....108.1906H}.
While type I is caused by magnetically induced cool spots corotating with the objects, type II
is related to the accretion flow and can contain periodic components, irregular eclipses,
and burst-like phenomena. In addition, young stars can show chromospheric flare events,
but we expect this to be rare in the near-infrared.

Type II variations are easily spotted by visual inspection of the lightcurves 
\citep[see example lightcurves in][]{2005A&A...429.1007S}. We examined the lightcurves of 
variable objects in the globule area carefully. This was restricted to the objects within
$<9'$ distance from IRAS 21246+5743. Out of 165 objects 89 fulfill this criterion.
Nine of them show well-defined variations after visual inspection, including two obvious
and regular eclipses (1-2750, 1-3677), discussed in Sect. \ref{seren}. None has 
the characteristic type II variations discussed above. 

In this sample of 89 objects we searched for periodicities. In most cases the variability
in YSOs is more pronounced in the J- than in the K-band \citep{2001AJ....121.3160C}; therefore the
period search is based on the J-band lightcurves. Taking into account our limited sample
of $<40$ datapoints per lightcurve, we refrain from using periodogram analysis, instead
we have chosen a robust and simple approach: For a set of periods ranging from 0.0 to 2.5\,d
in steps of 0.015\,d, a given lightcurve was fit with a sine function. After subtraction of 
the best-fitting sinecurve, our routine compares the variance in the original lightcurve 
with the variance in the residuals using the F-test (see Sect. \ref{var}). This essentially 
probes how much of the variability in the lightcurve can be attributed to a sinusoidal period. 

\begin{figure*}
\includegraphics[width=4cm,angle=-90]{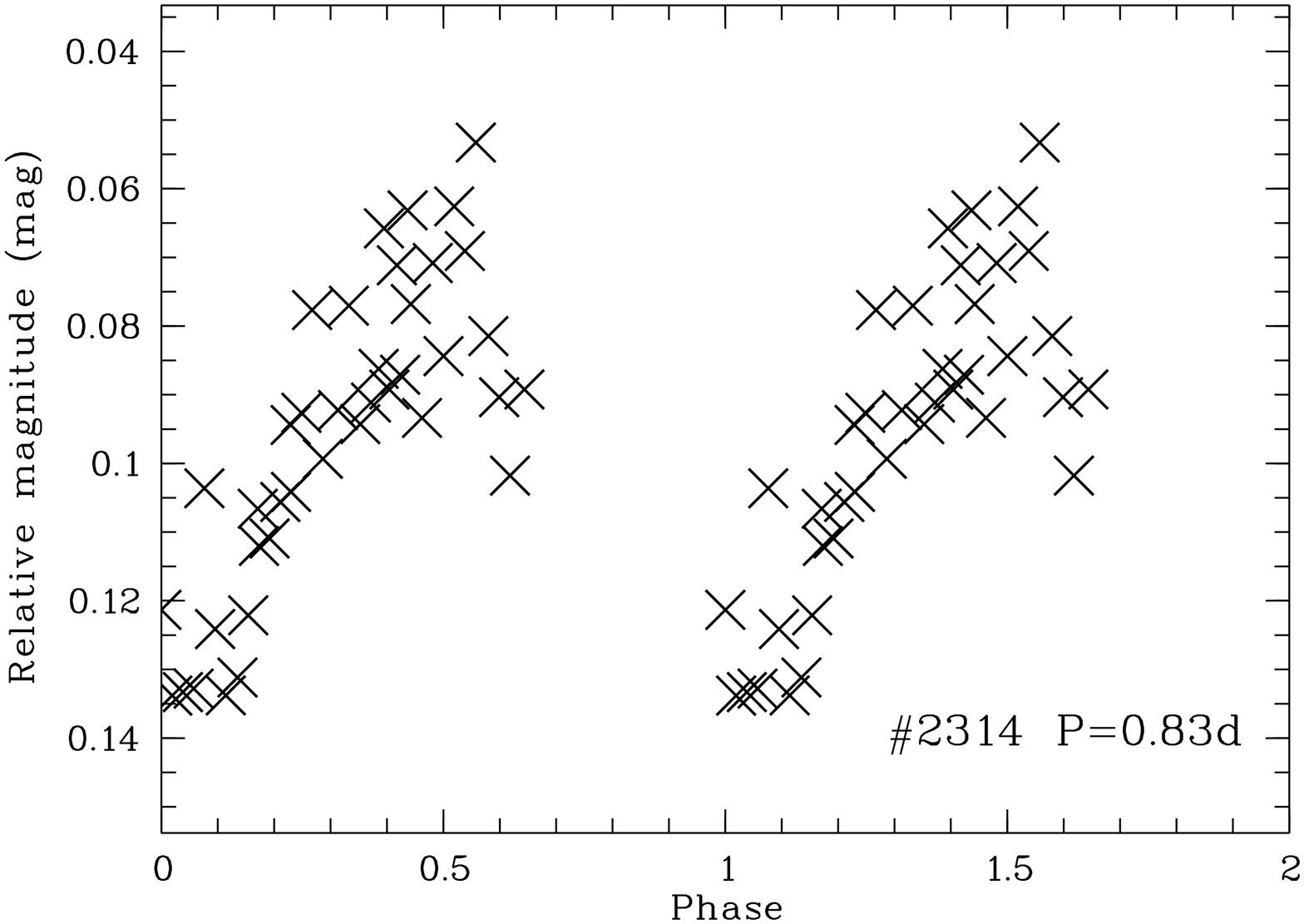} \hfill
\includegraphics[width=4cm,angle=-90]{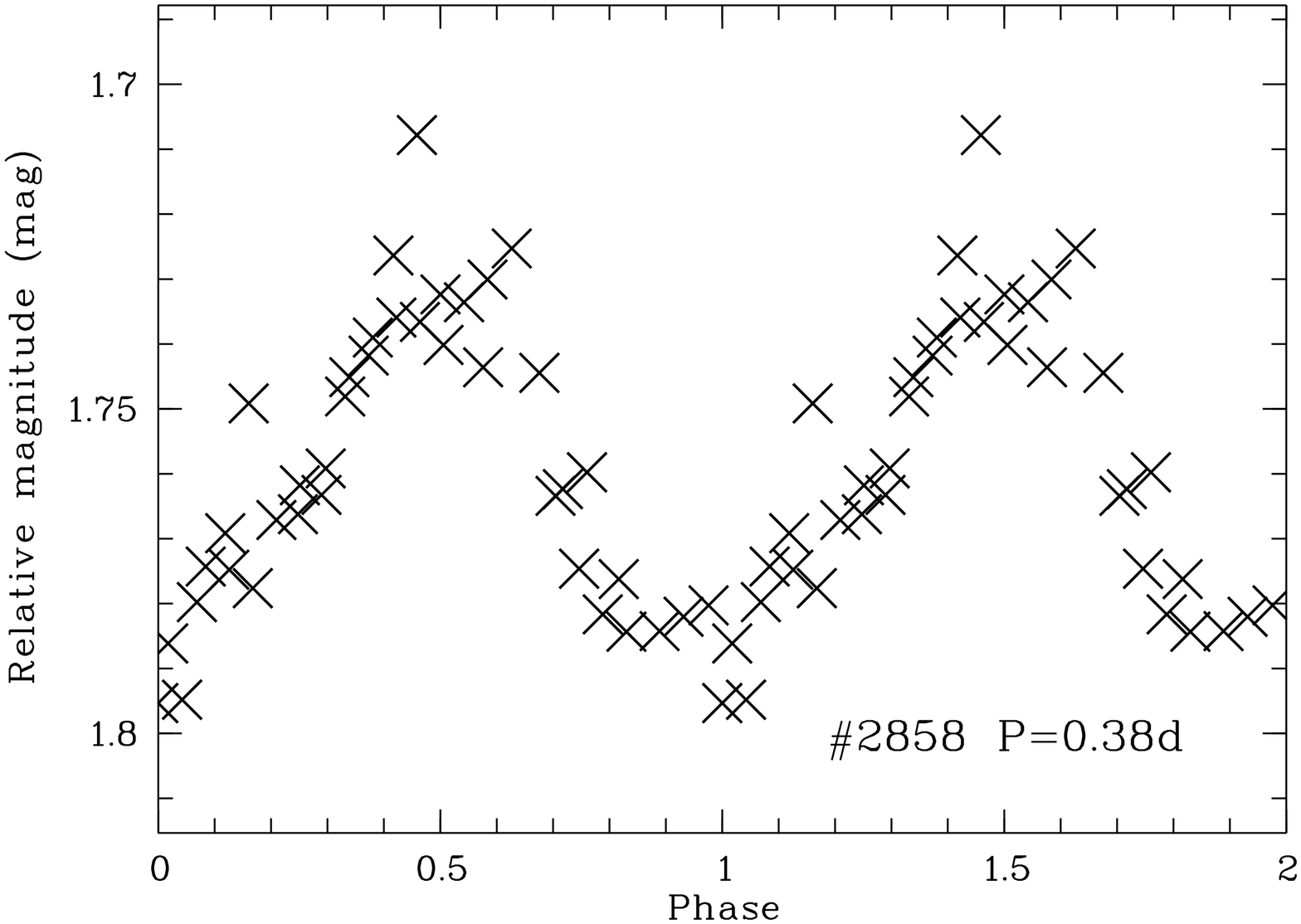} \hfill
\includegraphics[width=4cm,angle=-90]{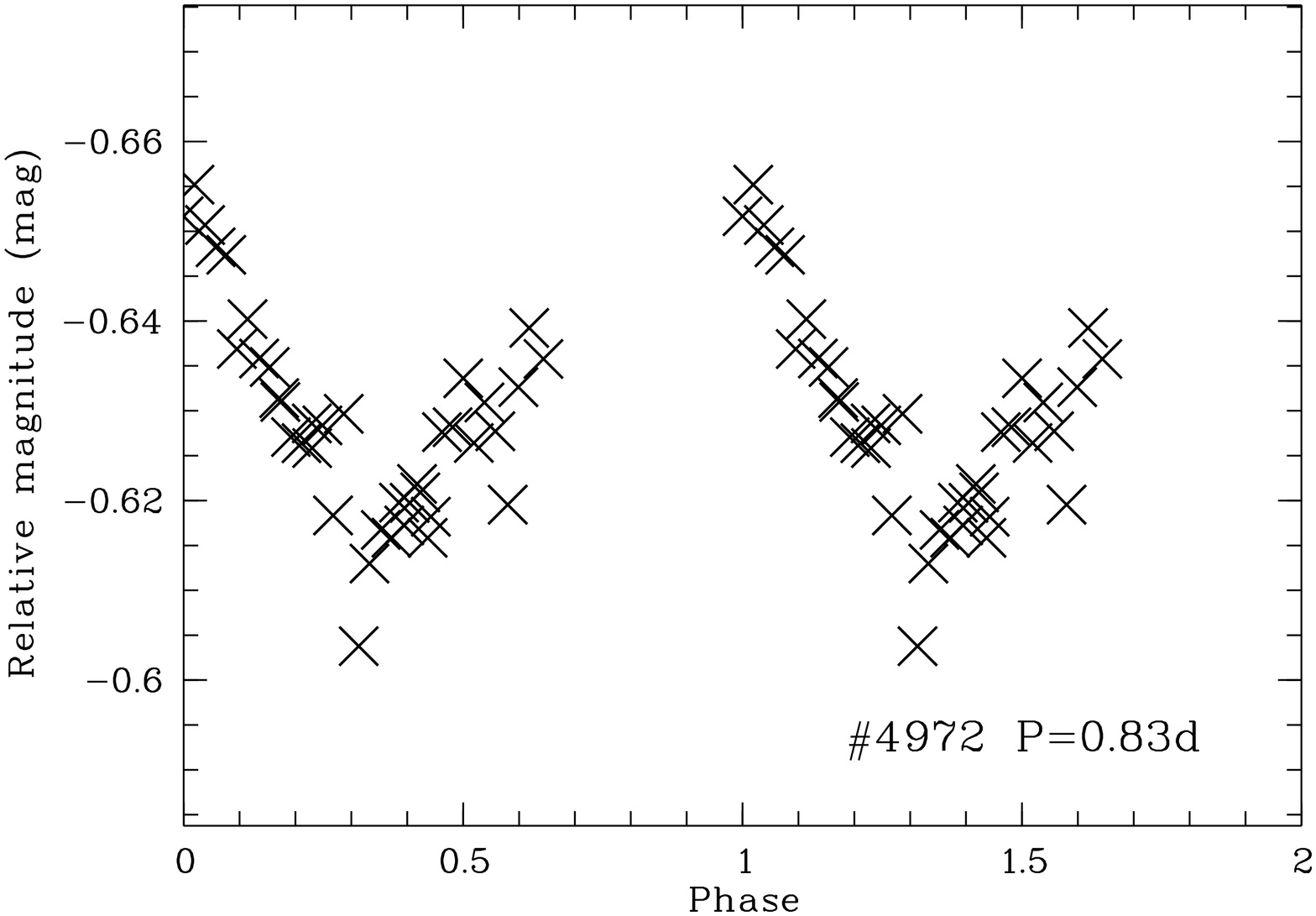} \\
\includegraphics[width=4cm,angle=-90]{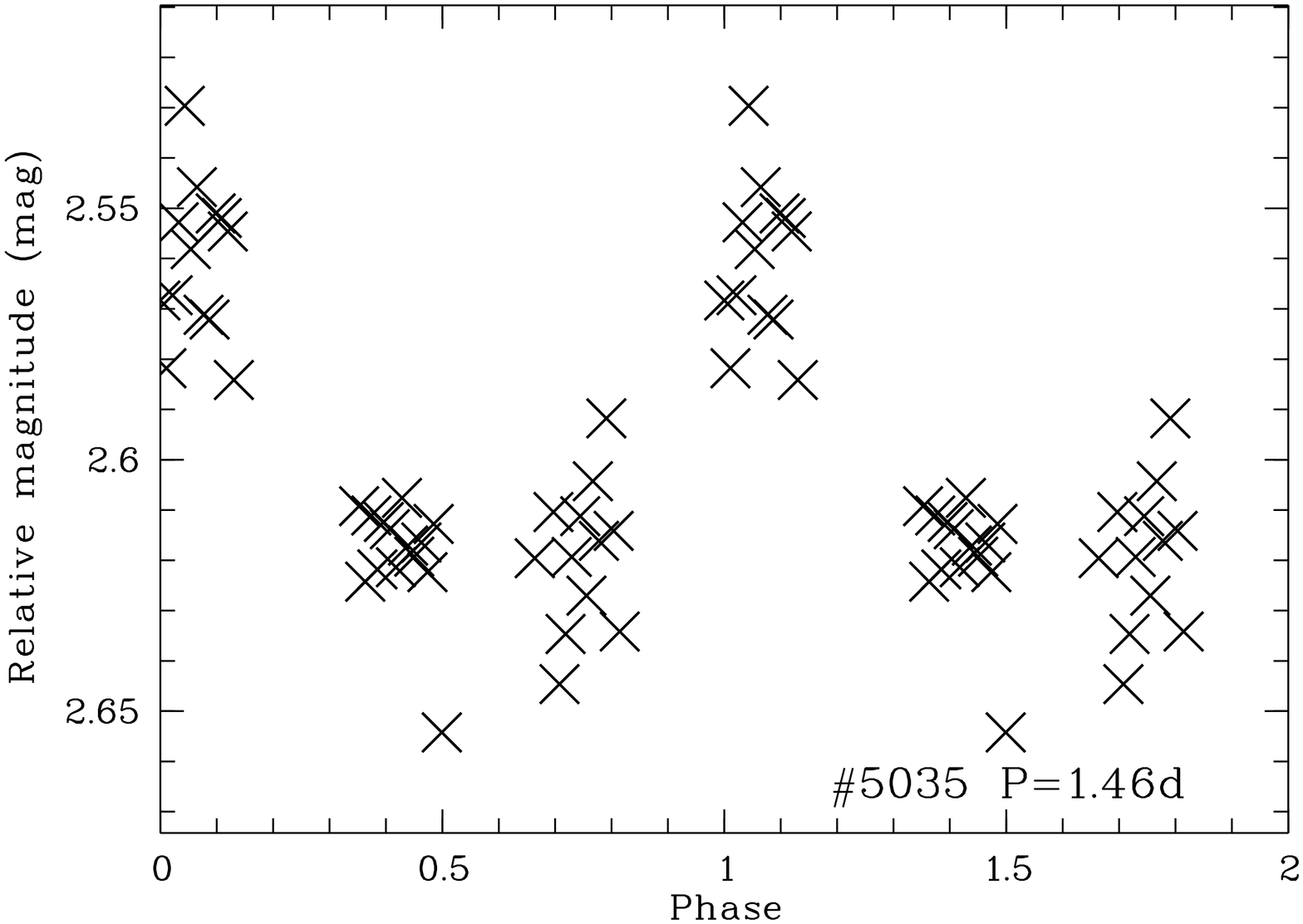} 
\includegraphics[width=4cm,angle=-90]{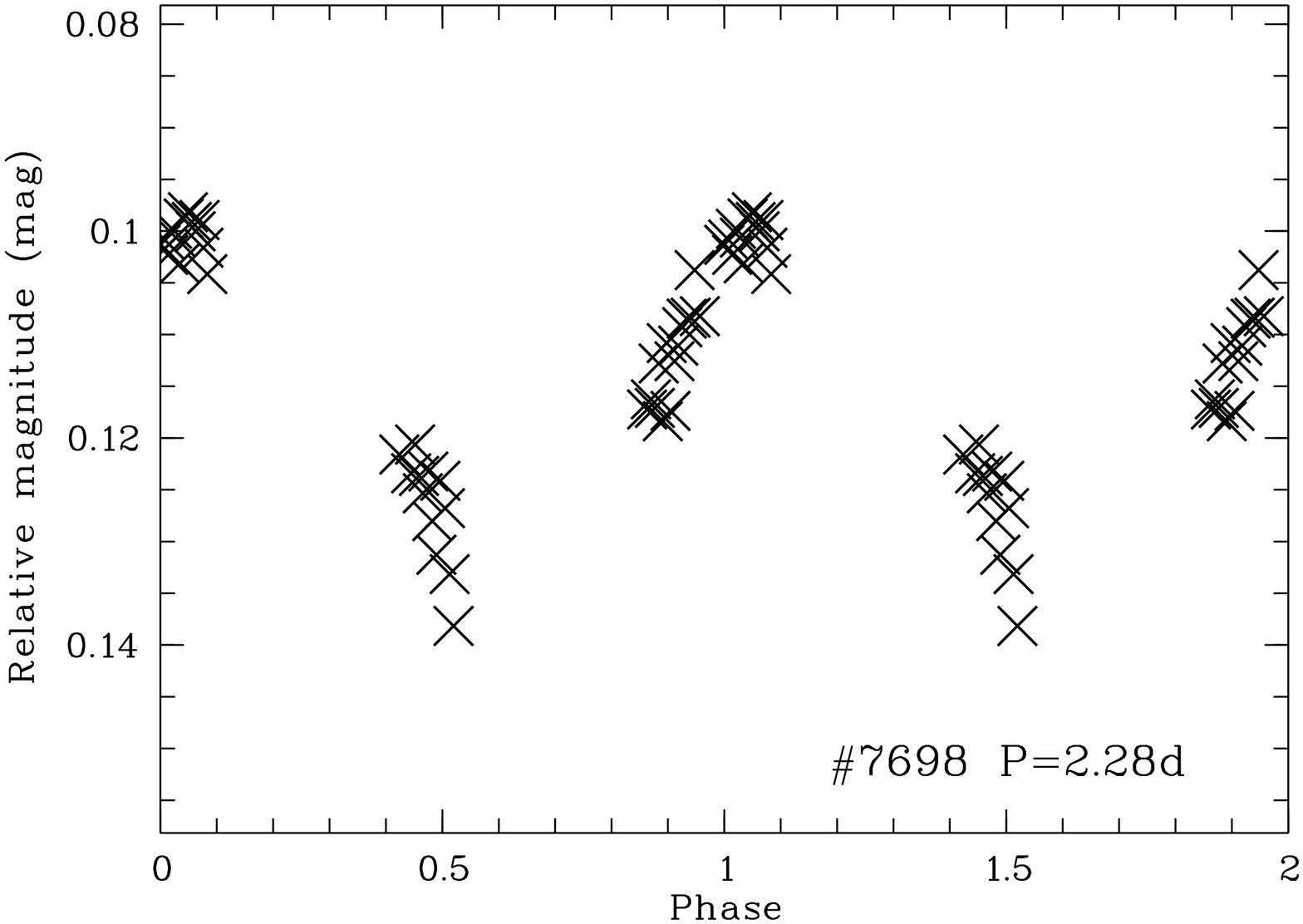} \\
\caption{J-band lightcurves for the stars with significant periodic variations plotted as function of phase.
Object ids refer to the catalogue for chip 1. \label{f4}}
\end{figure*}

Out of 89, 11 lightcurves exhibit a periodic signal with $F>2.5$, including the two eclipses.
Five objects have clearly pronounced variations with periods ranging from 0.4 to 2.3\,d. One 
of these objects the only one which fulfills the colour criterion in Sect. \ref{ccd}, is 1-2314,
located in the bright nebulosity discussed in Sect. \ref{comb}. We note that the second red and
variable object in this nebula (1-2189) shows a steady decline of 0.15\,mag over the three
observing nights, possible due to a longer period. The best five periods are shown in Fig. 
\ref{f4} (1-2314, 1-2858, 1-4972, 1-5035, 1-7698). All these periodicities have amplitudes 
$<0.1$\,mag. Amplitudes and periods are typical for rotational modulations in young stars, 
caused by cool spots. 

One more (1-6059) has a period of 0.25\,d and a large amplitude of 0.4\,mag. It has a J-band 
magnitude of 16.6, which would put it in the substellar regime, if it is a member of IC1396W, 
but the colour $H-K = 0.2$ is not matching. It is also relatively far away from the core of the 
globule (8'). This strongly argues against membership. This object is further discussed 
in Sect. \ref{seren}.

\subsection{H$\alpha$ photometry}
\label{halpha}

\begin{figure}
\includegraphics[width=6cm,angle=-90]{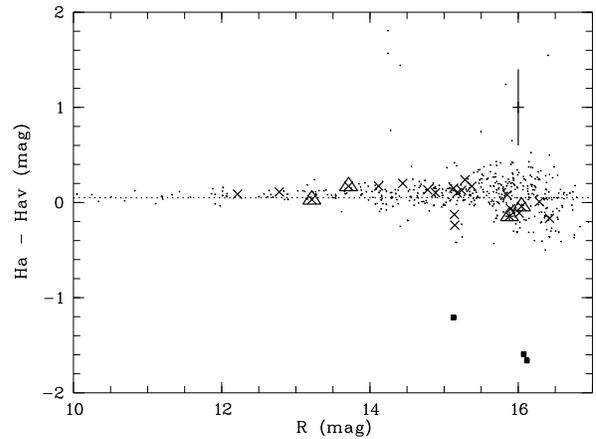} 
\caption{\label{f5} H$\alpha$ 'colour' vs. R-band magnitude for the IC1396W field, derived from the 
complementary optical observations. The filled squares mark objects with clear H$\alpha$ emission. 
Objects with $H-K$ colour excess are marked with plusses, variable objects with crosses, and objects 
with significant periods with triangles. The errorbarat $R=16$\,mag corresponds to the 3$\sigma$ 
uncertainty.}
\end{figure}

We use the optical photometry in the filter centered on the H$\alpha$ emission line as additional
check for young stellar objects in IC1396W. The optical catalogue contains 11357 objects with good
photometry in all bands, from which 533 are within a 9' radius of the globule center.
Fig. \ref{f5} shows the H$\alpha$ 'colour' (i.e. the difference in instrumental magnitudes 
between the flux in H$\alpha$ and the flux in a filter next to the H$\alpha$ feature) vs. 
R-band magnitude for these 533 objects. About 70\% of them have a counterpart in the
near-infrared catalogue.

For clusters with a strong YSO population this type of diagram is expected to show a bimodal
population (main-sequence and YSO), as for example seen in \citet{2004A&A...417..557L}. Our plot 
does not show any sign of bimodality, again indicating that IC1396W does not harbour a significant 
number of YSOs. The datapoints form a symmetric distribution around $H_a - H_{av} = 0.05$ which is 
likely the locus of the background main-sequence. The scatter of the datapoints at the faint 
end of the plot is consistent with the photometric uncertainty.

Objects with strong H$\alpha$ emission are expected to be located below this line. Only three 
objects fall below the line on a confidence level much better than 3$\sigma$. Two of them
are clearly identified in the K-band image as part of the SW head of the largest outflow in the cloud, 
which is visible in the optical bands as well. 1-2223 has a point-source counterpart in the near-infrared,
which is neither variable nor red; this might be a disk-less WTTS. A handful of additional objects
is close to the 3$\sigma$ limit; we do not consider them to be good candidates due to the lack
of separation in H$\alpha$ colour between them and the bulk of the datapoints.

Overplotted in Fig. \ref{f5} are red objects (plusses) according to our $H-K$ criterion on Sect. \ref{ccd} (only
object 1-2314, as most red sources are not detected in the optical), variable objects (crosses) in J- and K-band 
according to the criterion in Sect. \ref{var}, which includes four objects with significant periods 
(Sect. \ref{char}) shown with triangles.

\section{Star formation in IC1396W}
\label{yso}

\subsection{The YSO census}
\label{census}

Based on the discussion in Sect. \ref{ident} we will compile samples of candidate YSOs in
the IC1396W globule. We distinguish five groups of objects. For all five groups we additionally 
require the objects to be point sources and to be spatially located within the area of the 
globule ($<9'$ distance from IRAS 21246+5743):

\begin{enumerate}
\item{$H-K$ excess and variability: Only two objects fulfill these criteria,
with $K=11.3$ and $K=11.5$ (1-2189 and 1-2314). As mentioned in Sect. \ref{comb}, both 
are located close together in a bright nebulosity.}

\item{$H-K$ excess and located on the axis of one of the three outflows,
classifying them as possible outflow driving sources: One source with $K=10.7$ (1-5864) falls 
on the axis for the northwestern outflow. This objects has already been discussed in 
\citet{2003A&A...407..207F} as a possible driving source for this outflow.}

\item{$H-K$ excess and located below the reddening path in the colour-colour
diagram: From our C-sample, 78 objects are in the globule area, from which 54 look like
point sources. 7 of these objects are located significantly below the well-defined reddening
path of the remaining objects, with K-band magnitudes between 15.0 and 16.5. This sample 
likely contains reddenened background objects.}

\item{Variability characteristic typical for YSOs: From our V-sample, 89 objects are in
the globule area, and four of them show YSO-like variability (see Sect. \ref{char}), with
K-band magnitudes between 12.1 and 15.3. This sample likely contains variable field objects 
(e.g. active M dwarfs or short-period pulsating giants).}

\item{H$\alpha$ emission detected using our photometric criterion as defined in Sect. \ref{halpha}:
Only one object, 1-2223, fulfills this criterion, it has a K-band magnitude of 13.8.}
\end{enumerate}

In Table \ref{t2} we list the coordinates and photometry for the candidates selected
based on these four criteria; their positions are overplotted in Fig. \ref{f100}. In total, 15 
candidates are identified from colour, variability, and spatial position. All require 
spectroscopic confirmation, in particular the groups (iii)-(v). Five of them do not show 
colour excess and could thus be WTTS in IC1396W. 
In addition, the globule is known to harbour the likely Class 0 source 
IRAS 21246+5743 \citep{2003MNRAS.346..163F}. This can be contrasted with the 31 red objects 
identified in the shallow survey by \citet{2003A&A...407..207F}. A pure colour selection, 
as used by \citet{2003A&A...407..207F}, clearly overestimates the number of YSOs in the cloud.

The depth and completeness of our sample is not trivial to determine, because we used
a variety of different indicators. Our photometry database requires the objects to have
uncertainties $<0.2$\,mag in K- and J-band, which effectively limits the survey to
$J<19$\,mag (Sect. \ref{ccd}). Assuming a maximum extinction of $A_V=10$\,mag and a distance
of 750\,pc, this corresponds to $M_J \sim 7$. With the 1\,Myr track by 
\citet{1998A&A...337..403B} this yields a mass limit $\sim 0.05\,M_{\odot}$, which
is applicable to categories (i), (ii) and (iii) above. Since all objects identified
in category (iv) have amplitudes $<0.1$\,mag, the depth for this sample is likely
to be 2-3\,mag lower, corresponding to $\sim 0.2\,M_{\odot}$. Thus, the survey covers
the regime around the peak in the IMF. 

The sample of candidate YSOs in the categories (iii)-(v) may contain a substantial fraction of
contaminating background objects. On the other hand our method will miss certain types of sources 
(see discussion in \ref{ident}). It is sensitive to CTTS and variable WTTS, which are the major 
fractions of objects in regions at 1\,Myr. Therefore we do not expect that the incompleteness affects
our most relevant result: The number of candidate YSOs in this globule is found to be low, 
probably less than 10.

\begin{table}
\centering
\caption{Candidate YSOs in IC1396W. The five groups of objects in this table
correspond to the groups (i)-(v) described in the text (Sect. \ref{census}).
\label{t2}}
\begin{tabular}{rcccccr}
\hline
ID & $\alpha$ (J2000) & $\delta$ (J2000) & $K$ (mag) & $H-K$ & $J-H$\\
\hline
1-2189 & 21:26:31.44 & 57:55:50.90 & 11.29 & 1.00 & 1.85\\
1-2314 & 21:26:31.53 & 57:55:52.80 & 11.51 & 0.76 & 1.58\\
\hline
1-5864 & 21:25:41.63 & 57:57:18.6 & 10.68 & 1.40 & 3.15\footnotemark \\ 
\hline 
1-2368 & 21:26:31.30 & 57:54:05.2 & 15.53 & 0.77 & 1.26 \\ 
1-2560 & 21:26:29.12 & 57:52:45.1 & 15.75 & 1.18 & 1.30 \\ 
1-3314 & 21:26:19.58 & 58:01:02.6 & 15.80 & 0.79 & 1.44 \\ 
1-3366 & 21:26:19.16 & 58:00:51.7 & 16.14 & 0.89 & 1.64 \\ 
1-4467 & 21:26:03.49 & 57:54:09.6 & 14.96 & 0.97 & 1.60 \\ 
1-5828 & 21:25:46.53 & 57:51:02.7 & 16.48 & 0.99 & 0.95 \\ 
1-7003 & 21:25:30.81 & 57:54:52.2 & 16.50 & 1.07 & 1.46 \\ 
\hline
1-2858 & 21:26:25.26 & 57:55:59.5 & 14.40 & 0.24 & 0.87 \\ 
1-4972 & 21:25:56.16 & 58:02:16.9 & 12.11 & 0.27 & 0.74 \\ 
1-5035 & 21:25:56.98 & 57:50:26.3 & 15.33 & 0.24 & 0.77 \\ 
1-7698 & 21:25:38.58 & 57:54:14.6 & 13.08 & 0.18 & 0.59 \\ 
\hline
1-2223 & 21 26 31.57 & 57 55 52.6 & 13.84 & 0.25 & 0.82 \\ 
\hline
\end{tabular}
$^1$ possible source for outflow 3, see \citet{2003A&A...407..207F}
\end{table}

\subsection{Star formation efficiency}
\label{sfe}

IC1396W is one of the largest and most massive clouds in the IC1396 region. \citet{2005A&A...432..575F}
determined a globule mass of 400-550$\,M_{\odot}$ from near-infrared extinction maps, assuming a distance
of 750\,pc. Combined with the low number of YSOs this indicates low star formation efficiency. Before we 
can quantify this more accurately, however, we re-assess the distance of IC1396W.

A standard way of probing distances to dense clouds is by separating foreground and background objects
using the near-infrared colour and comparing the number of foreground objects with predictions from stellar
population models. In IC1396W this procedure is only applicable to the innermost part of the cloud, where 
the extinction is strong enough to block the background objects. Within 1.5' from
the IRAS source there are 8 objects with unreddened colours of $J-K \sim 1.0$, which agrees well with the
typical colours of objects outside the cloud. The other 14 objects in this area are clearly reddened
with $J-K>1.5$. 

We used the Galaxy model from the Besancon group\footnote{http://model.obs-besancon.fr/} to simulate a
catalogue of objects over an area of 1\,sqdeg in the direction of IC1396W matching the dynamic range
of our catalogue ($9.3<K<18.7$\,mag). This yields 4798 objects with distance $<750$\,pc, which corresponds
to 9 objects for an area of $\pi R^2$ with $R=1.5'$. Thus, the model predicts a number of foreground
objects that is remarkably close to the actual number of foreground objects. This indicates that the
assumption of $d=750$\,pc is plausible. Assuming Poissonian errors, the distance is unlikely to be below
$d=600$\,pc. 

Thus the estimates for the cloud mass reported above are valid. As discussed in Sect. \ref{census} the
number of YSOs in IC1396W is likely to be around 10 or less. Assuming an average mass of $\sim 0.5\,M_{\odot}$
gives a total stellar mass of 5$\,M_{\odot}$ and a star forming efficiency SFE around 1\%. This can be
compared with SFE values determined for nearby star forming regions: According
to \citet{2009ApJS..181..321E} star formation is significantly more efficient ($\ge 3$\%) in Cha\,II, Lupus, 
Perseus, Serpens, and Ophiuchus. With the exception of Cha\,II, these clouds are much more massive than
IC1396W. On the other hand, IC1396W is comparable in SFE with cores without stellar clusters in Cepheus 
\citep{2009ApJS..185..198K}, but exceeds all these regions in mass. Thus, our survey establishes
IC1396W as an intermediate case of a relatively massive cloud with low star formation efficiency.
The SFE is still about one order of magnitude higher than in the Pipe nebula
\citep[0.06\%,][]{2009ApJ...704..292F}, which appears to be a cloud in the earliest phases of 
star formation.

There are indications that the radiation from the O6.5V star in the center of the HII region IC1396
has an impact on the star forming activity in the surrounding clouds \citep{2005A&A...432..575F},
i.e. clouds at larger distances are bigger and less active. IC1396W is one of the more remote clouds in this
region, the distance from the exciting bright star is around 25\,pc. This could explain why the star
forming efficiency in IC1396W is found to be low.
 
\section{Serendipitous discoveries}
\label{seren}

While visually inspecting the lightcurves, we noticed a number of obvious variable objects.
In the following we report on the most prominent examples, including two eclipsing binaries
and 8 periodic variables. All these objects fulfill the conditions for variable objects
as outlined in Sect. \ref{var}. For the eclipsing binaries we present additional low-resolution 
spectra. Further analysis of these serendipitously discovered objects is postponed for future
studies.

\subsection{Eclipsing binaries}

\begin{figure*}
\includegraphics[width=6cm,angle=-90]{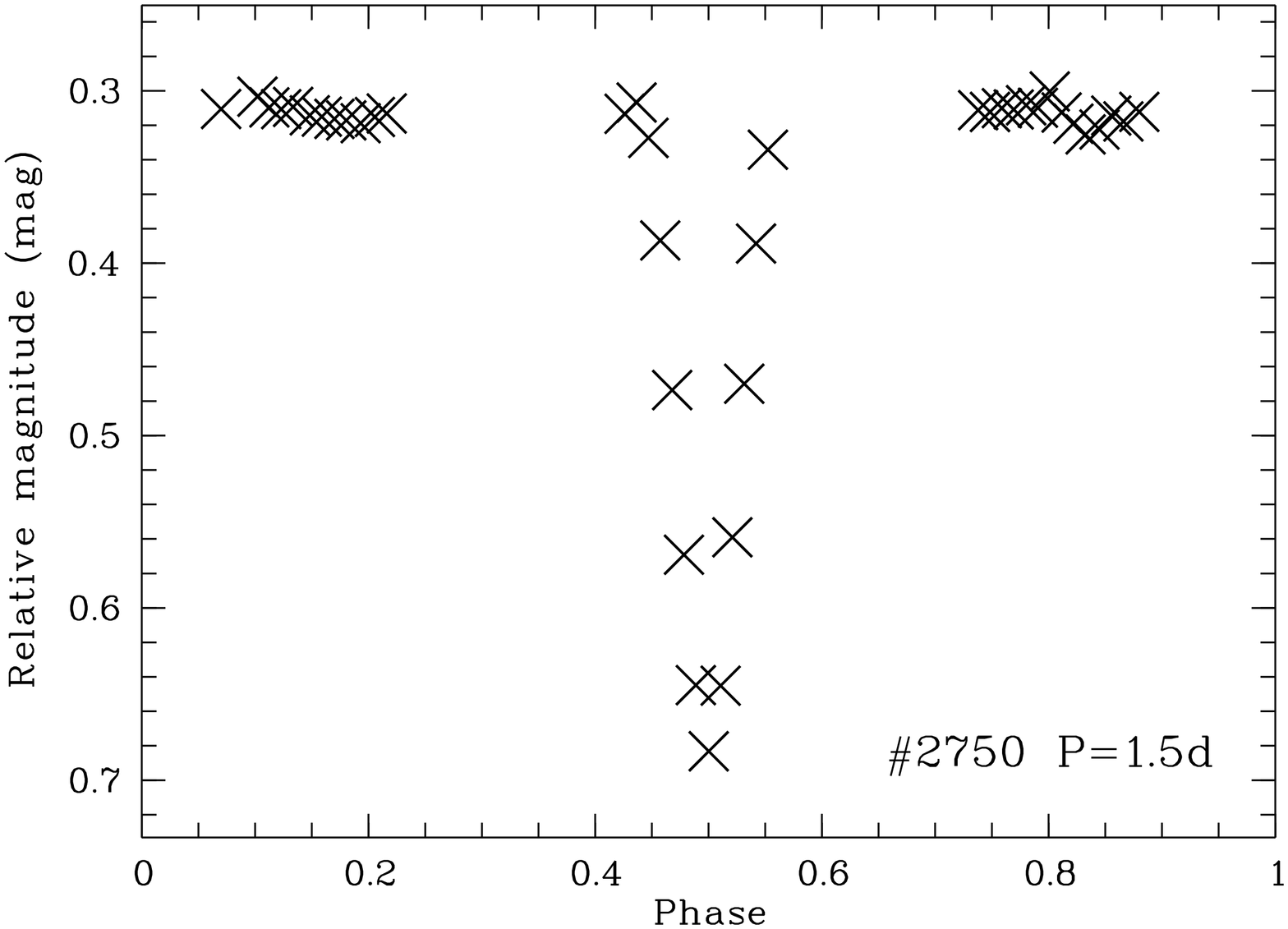} \hfill
\includegraphics[width=6cm,angle=-90]{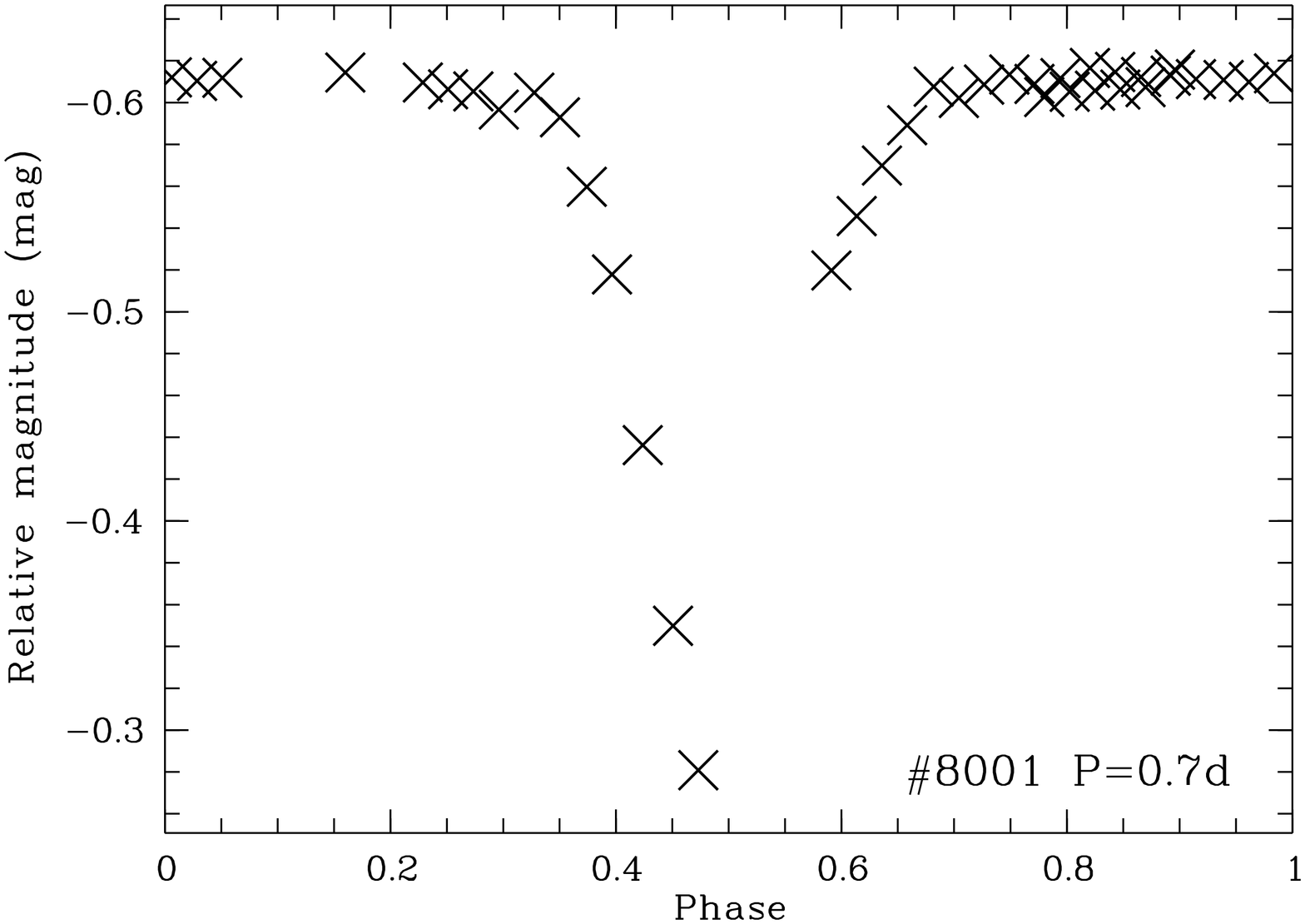} \\
\caption{Eclipses found in our lightcurves, J-band datapoints plotted as function of phase. The periods are
not constrained by our lightcurves; shown are examplarily fits. Object ids refer to the catalogue for 
chip 1. \label{f6}}
\end{figure*}

In the field of IC1396W there are two stars which show clear, deep eclipses in our lightcurves.
In both cases we observe only one eclipse, i.e. we cannot constrain the period. Based on the
symmetry of the ellipses and the smooth ingress/egress, we interpret the sources as eclipsing
binaries, although in principle other origins are conceivable (e.g., eclipses by circumstellar
material). The parameters
of these systems are given in Table \ref{t3}, their lightcurves are plotted in Fig. \ref{f6}.

\begin{table*}
\centering
\caption{Eclipsing binaries in our monitoring field. \label{t3}}
\begin{tabular}{rcccccrr}
\hline
ID & $\alpha$ (J2000) & $\delta$ (J2000) & $K$ (mag) & $H-K$ & $J-H$ & $\Delta J$ & $\Delta K$ \\
\hline
1-2750 & 21:26:25.71 & 57:59:02.4 & 12.85 & 0.35 & 0.84 & 0.38    & 0.37    \\ 
1-8001 & 21:25:23.40 & 57:49:35.5 & 12.42 & 0.24 & 0.49 & $>0.06$ & $>0.07$ \\ 
\hline
\end{tabular}
\end{table*}

\begin{figure}
\includegraphics[width=6cm,angle=-90]{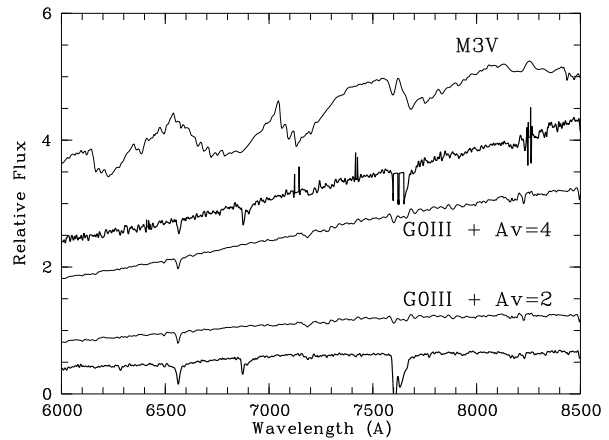}
\caption{Low-resolution spectra for the two eclipsing binaries near the cloud IC1396W. The thick
solid lines are the spectra for 1-2750 (upper) and 1-8001 (lower). The thin lines show template
spectra for a reddened G0III giant and a M3V dwarf from \citet{1998PASP..110..863P}. The absorption features at 
6850 and 7600\,\AA\,are telluric bands. The spikes at 7100 and 7400\,\AA\,in the spectrum of 1-2750
are caused by cosmics.\label{f8}}
\end{figure}

If these eclipsing binaries are associated with the young population of IC1396W, they are 
of particular interest to constrain the fundamental properties of young low-mass objects. This
motivated us to obtain follow-up spectra to look for evidence of youth. The ISIS spectrograph
at the William Herschel Telescope was used to observe these objects with grisms R158R (1.8\,\AA\,per
pixel, 30\,min integration time, date 18/07/2008) and R1200R (0.26\,\AA\,per pixel, 30\,min, 21/07/2007
and 13/10/2008). All data was taken in Service Mode. A spectroscopic standard reduction was carried 
out for these spectra, including bias and flatfield correction and background removal by subtracting a 
onedimensional fit along the spatial direction. The spectra are exctracted using {\tt apall} within IRAF.
Wavelength calibration is done based on Cu-Ar arc spectra. The low-resolution data
is flux calibrated based on a spectrum for the standard star SP2157+261. The results
are shown in Fig. \ref{f8}.

Based on the spectra it is safe to say that the eclipsing binaries are not associated
with the young population in IC1396W. Three arguments are important here: a) There is no
evidence for H$\alpha$ emission typical for YSOs, either due to accretion or magnetic
activity. In fact, H$\alpha$ is observed as an unshifted absorption line which also excludes 
the interpretation as extragalactic objects. b) The high-resolution
spectra, albeit at low signal-to-noise ratio, do not show any evidence for the deep
Lithium absorption feature at 6709\AA\,which is found in young stars with equivalent widths 
of $\sim 0.5$\,\AA\,\citep{2008ApJ...689.1127M}. c) If the objects are at the
distance of IC1396W and have an age of 1\,Myr they are expected to be M-type stars
\citep{2003ApJ...593.1093L}, which is inconsistent with the spectra (see Fig. \ref{f8}).
In particular, the characteristic TiO bandhead at 7050\,\AA\,and the CaH absorption at 
6830\,\AA\,are absent and the slope is almost flat. 

Additionally, the two objects are $>5'$ away from the IRAS source in the core of the cloud. 
Their colours indicate little reddening and there is no significant colour variability. All 
this argues against association with IC1396W. We conclude that these are objects in the background 
of the cloud.

The spectra are well-approximated with F- or G-type templates with little reddening,
as seen in Fig. \ref{f8}. This is also consistent with the near-infrared colours: For a G0-type 
star (luminosity class III or V) the intrinsic $H-K$ is 0.05 \citep{1988PASP..100.1134B}, 
which yields $H-K=0.2$ for $A_V=2$ (E-2) and 0.35 for $A_V=4$\,mag (E-1). The same testcase 
gives $J-H=0.6$ for $A_V=2$ and 0.85 for $A_V=4$. All this is in perfect agreement with our 
measurements. 

From the colours and the low-resolution spectra we cannot definitely tell if these two 
are giants or dwarfs. For a G0 dwarf the K-band flux would indicate a distance of
$\sim 1$\,kpc, for a giant about 5\,kpc. The sharpness of the eclipses favours dwarfs, 
as a giant is more likely to produce a slow ingress and egress. Another argument for
dwarfs is the relatively low extinction. (With $A_V \sim 0.7$\,mag/kpc we would expect
$H_K>0.5$ for giants.) 

The lack of a flat phase 
in eclipse, the depth of the eclipses, and the lack of colour change in eclipse argues for 
systems with roughly equal size components. Due to their faintness we did not attempt to 
obtain radial velocity monitoring to constrain their fundamental parameters. Anybody 
interested in working on these objects is encouraged to get in touch with us.

\subsection{Periodic variables}

\begin{figure*}
\includegraphics[width=4cm,angle=-90]{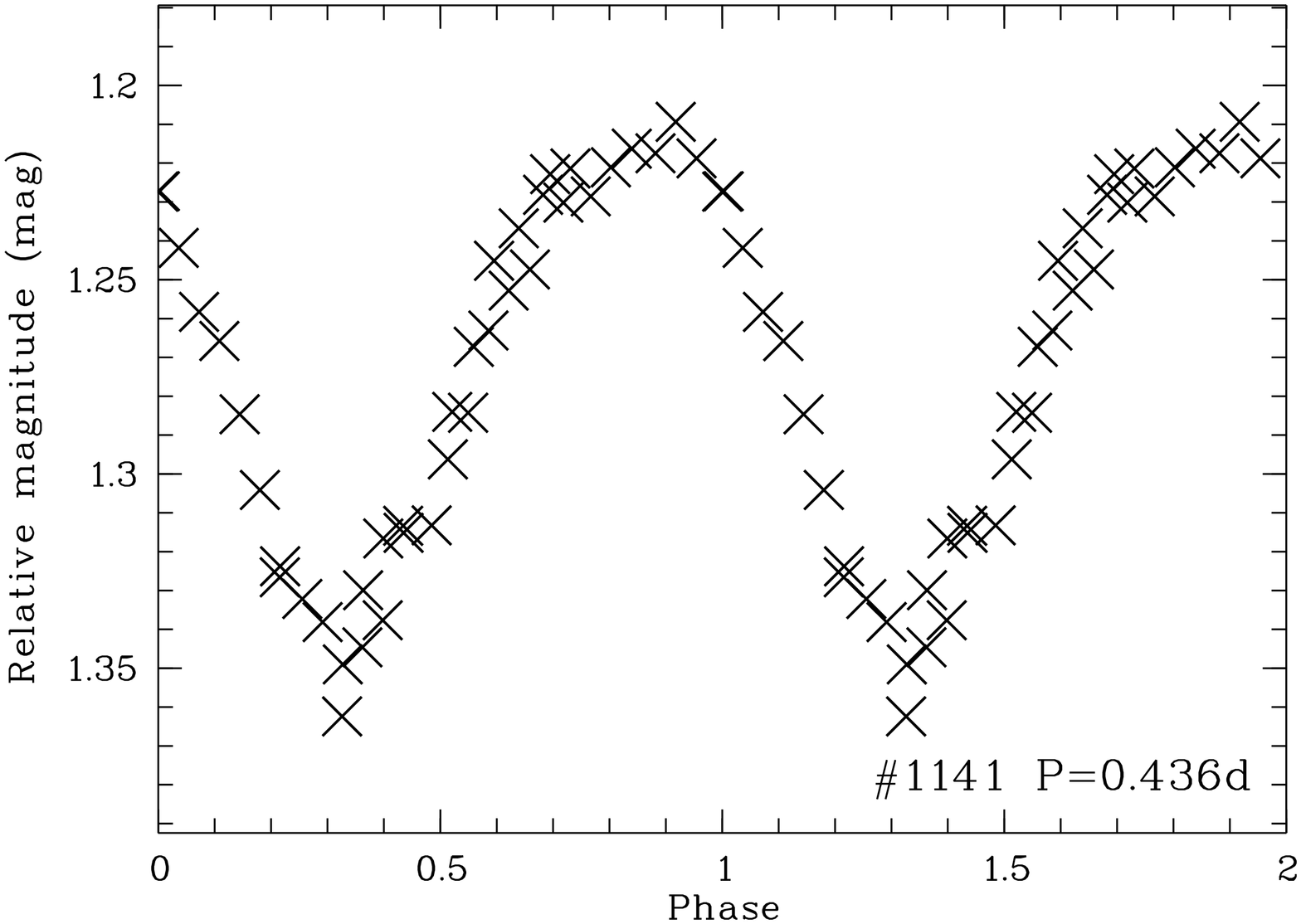} \hfill
\includegraphics[width=4cm,angle=-90]{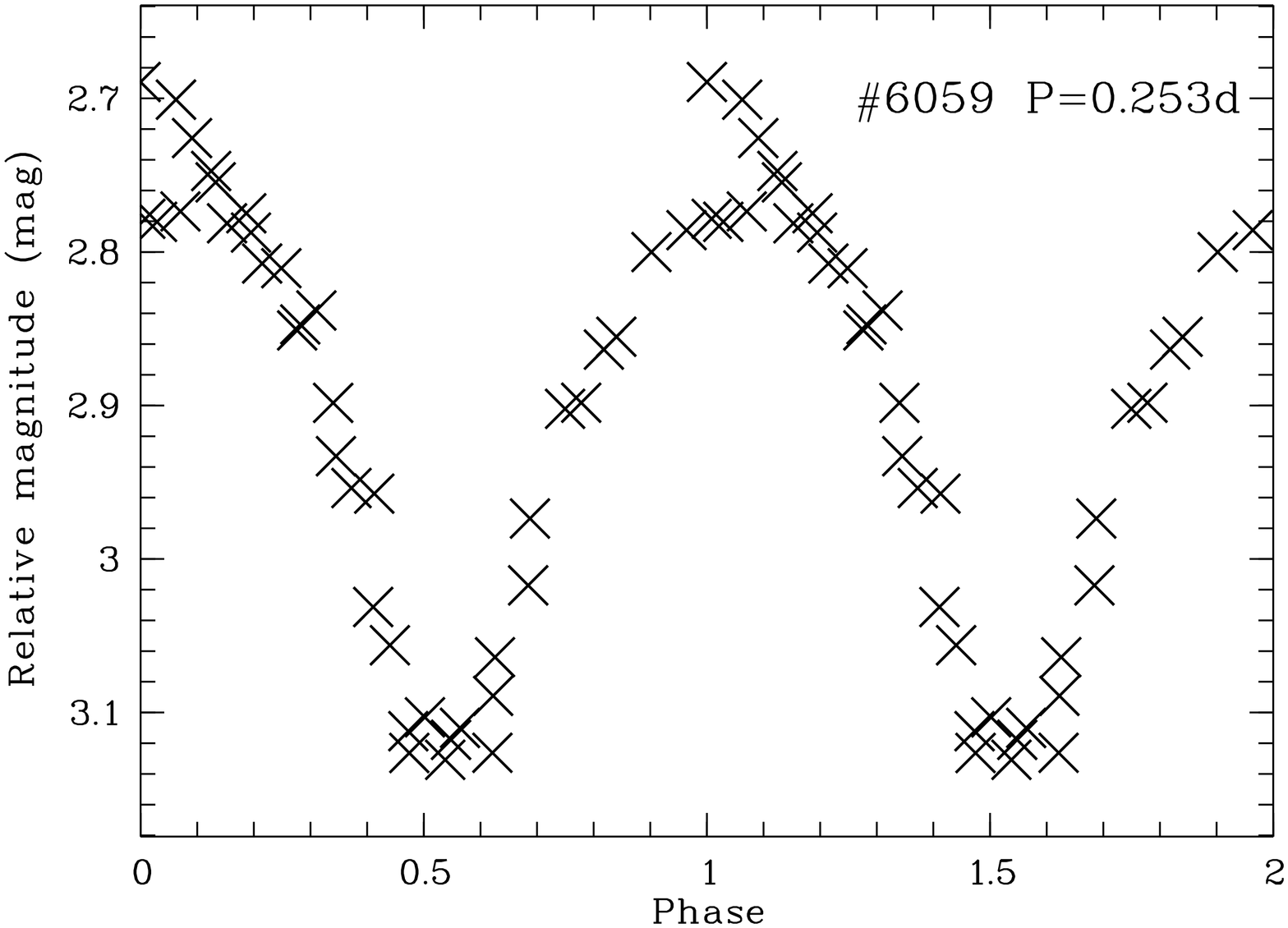} \hfill
\includegraphics[width=4cm,angle=-90]{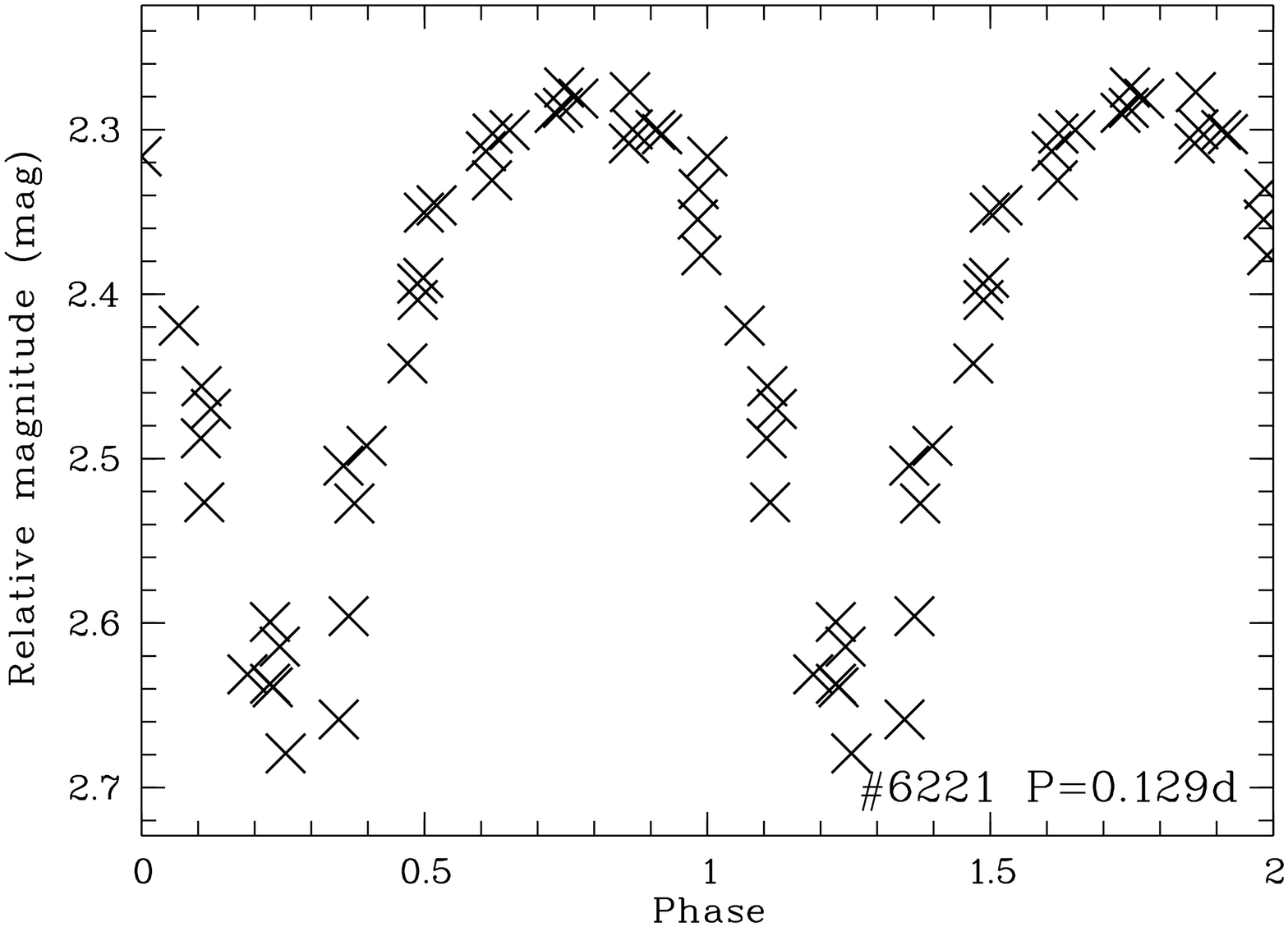} \\
\includegraphics[width=4cm,angle=-90]{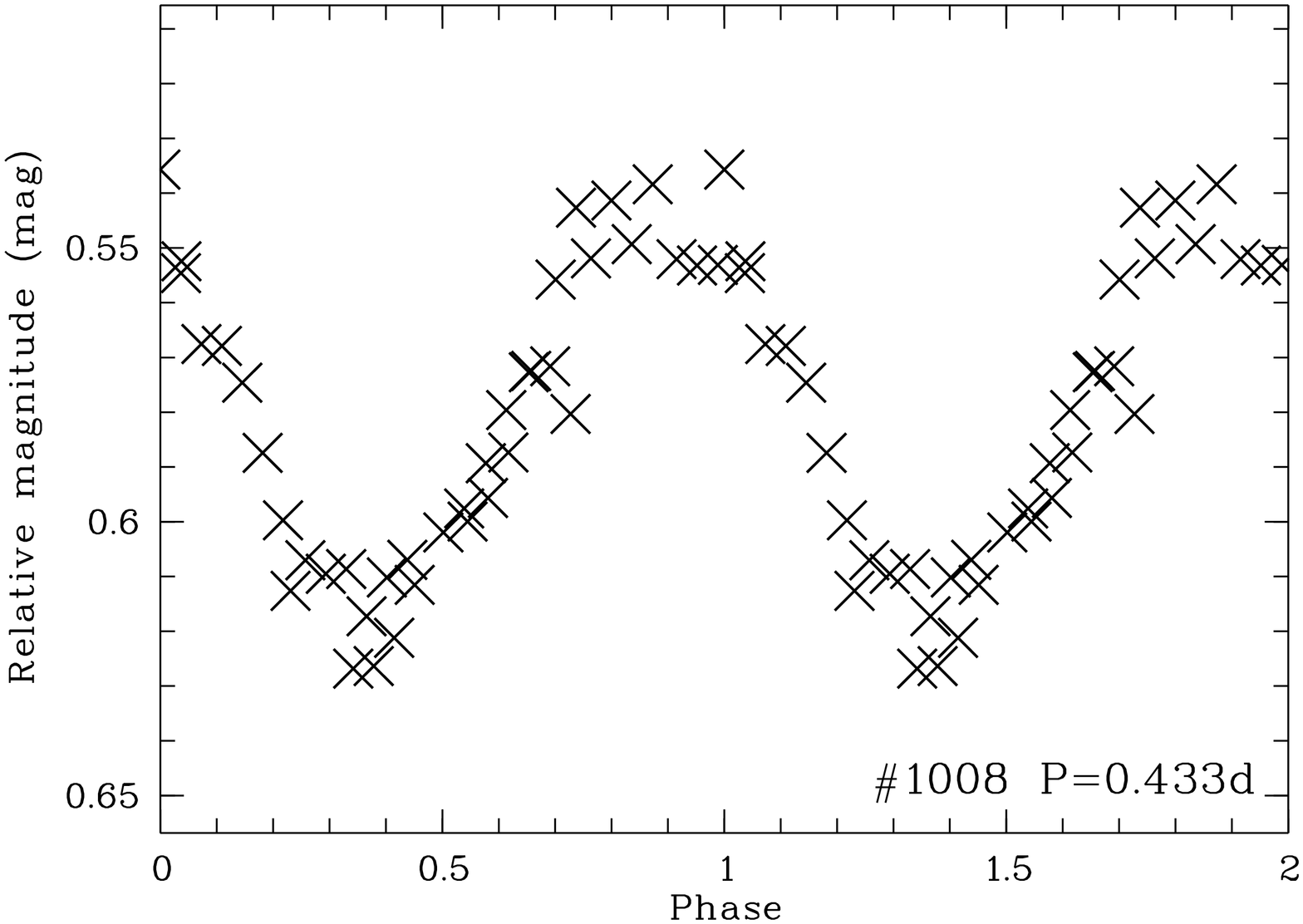} \hfill
\includegraphics[width=4cm,angle=-90]{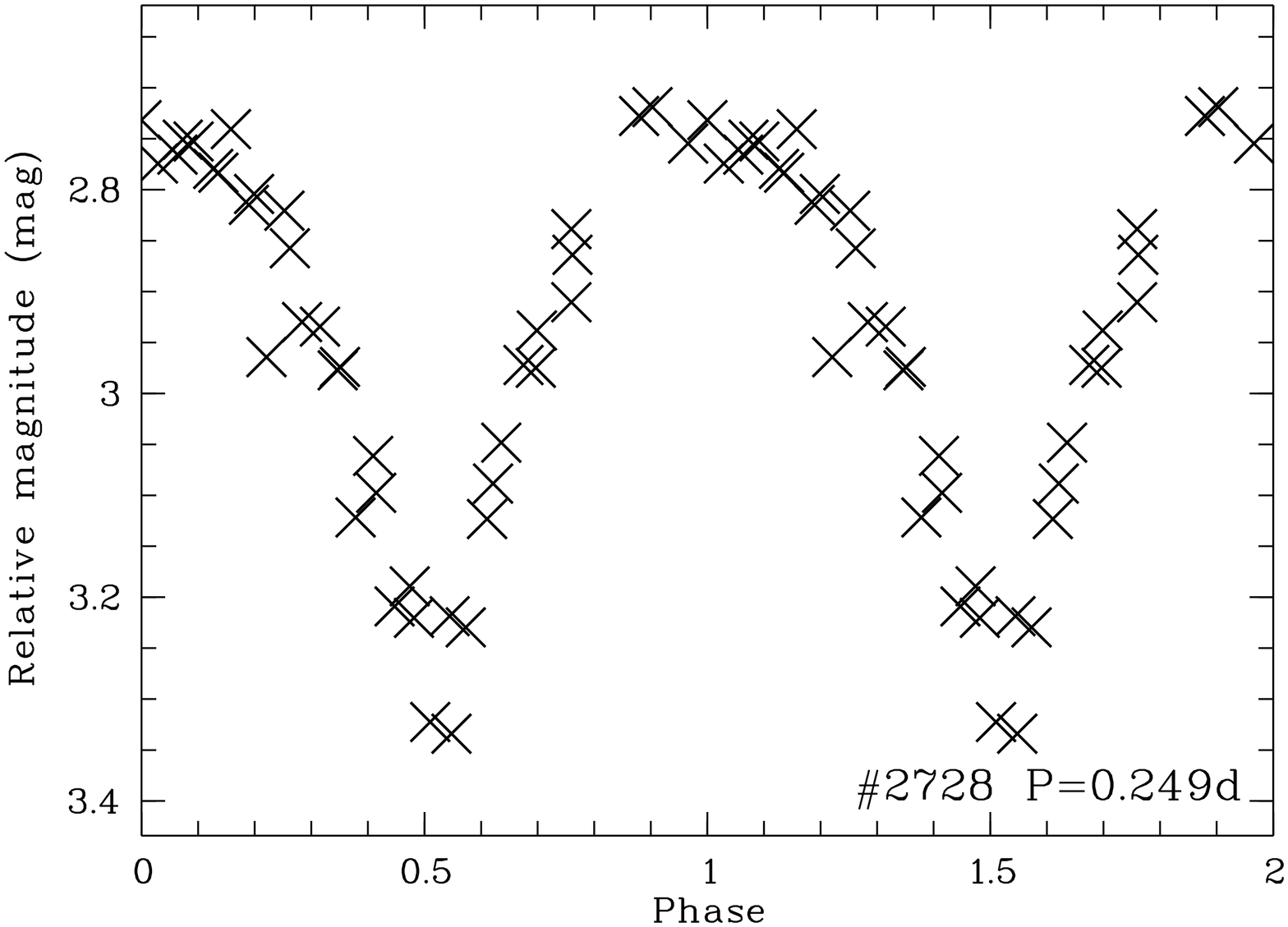} \hfill
\includegraphics[width=4cm,angle=-90]{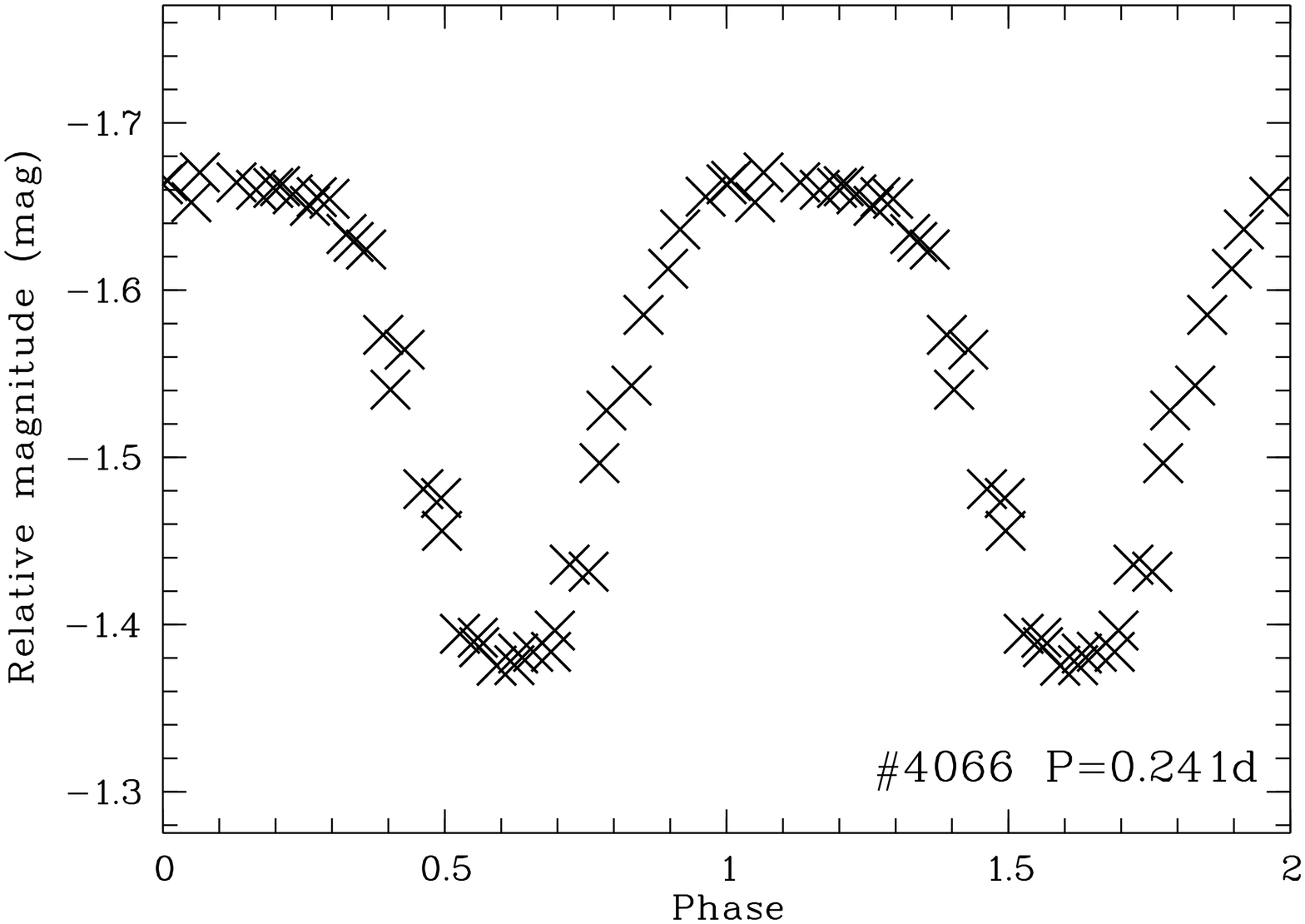} \\
\includegraphics[width=4cm,angle=-90]{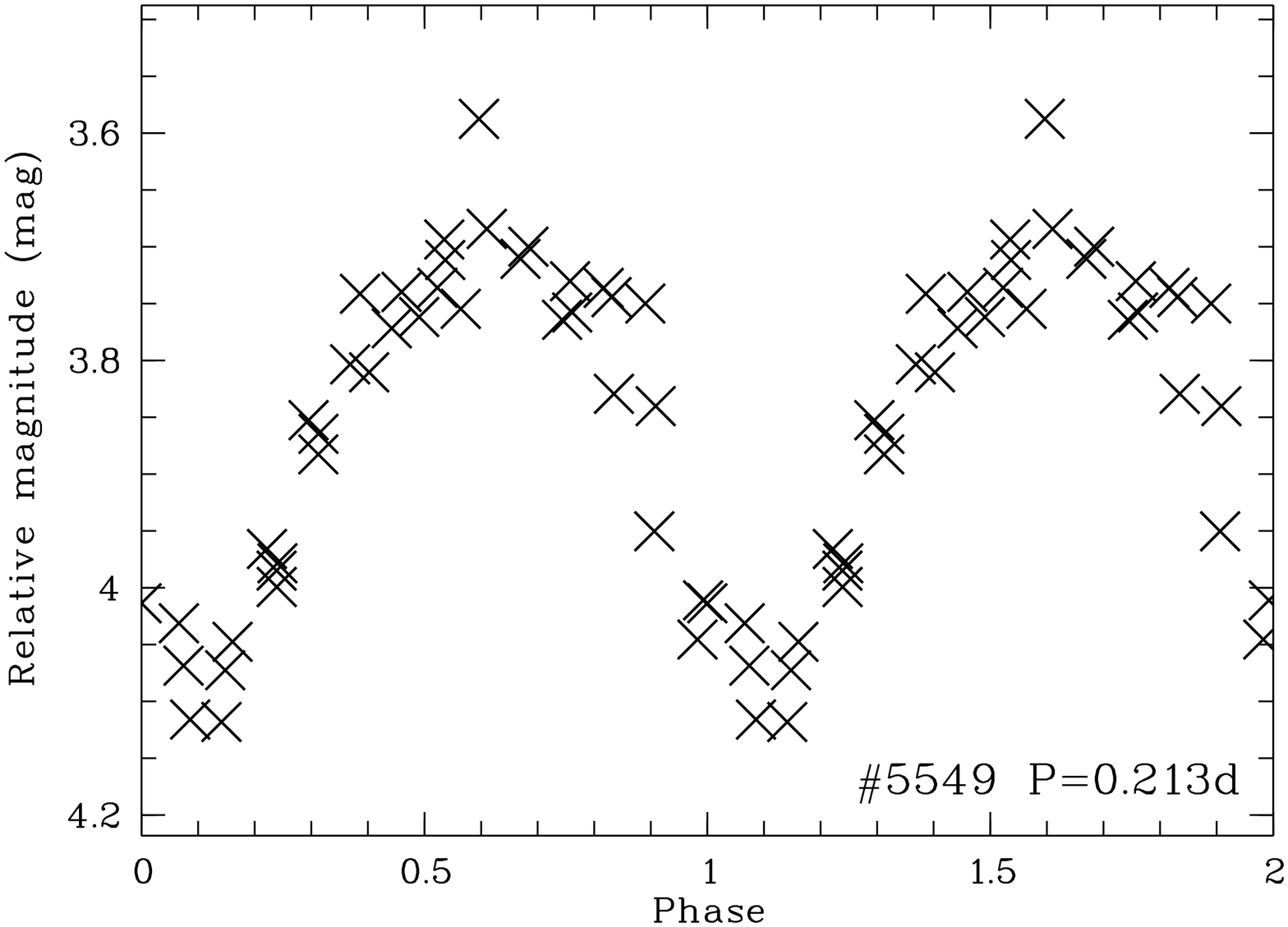} 
\includegraphics[width=4cm,angle=-90]{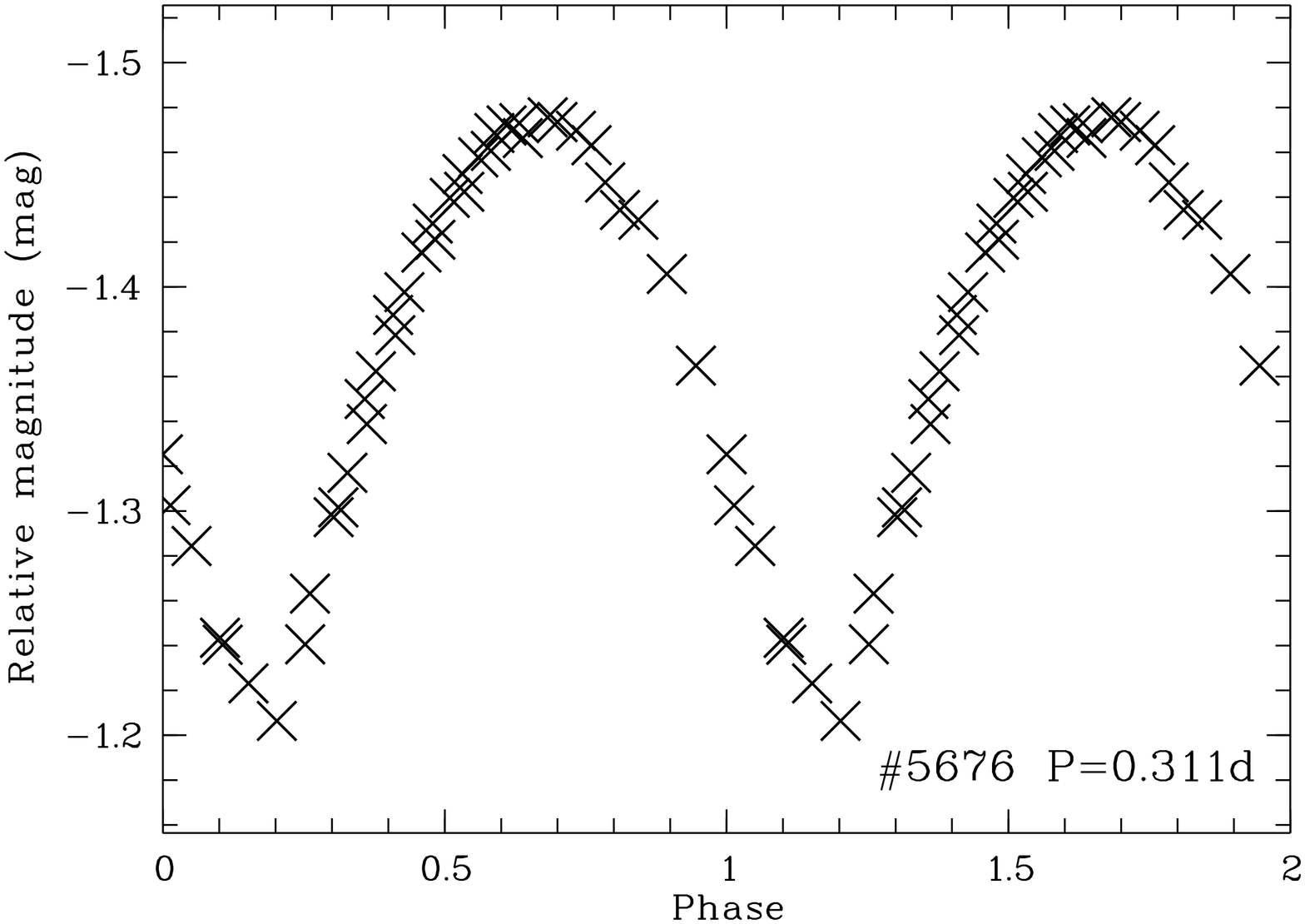}
\caption{Periodic variables found in our lightcurves, J-band datapoints plotted as function of phase. 
Object ids refer to the catalogue for chip 1 for the 1st row of panels, to the catalogue for chip 4
for the 2nd and 3rd row. \label{f7}}
\end{figure*}

In our database for the IC1396W field and the comparison field we noticed 9 obvious
short-period variables with large amplitudes (see Table \ref{t4}). The periods for these 
objects were estimated using the procedure outlined in Sect. \ref{char}. They all have periods 
ranging from 0.1 to 0.4\,d and J-band amplitudes from 0.1 to 0.6\,mag. We show their
lightcurves plotted in phase to the period in Fig. \ref{f7}.
Most of the lightcurves show evidence for changes in the brightness of maxima and minima on 
timescales of a few cycles, which is best explained by multiple periodicities being superimposed 
in the lightcurve. 

Lacking spectra it 
is difficult to definitely determine the nature of these objects. Most of them are good 
candidates for contact binaries. The CB frequency relative to field stars has been determined 
to $1/500$ for $M_V>+1.5$ \citep{2002PASP..114.1124R}. Both fields combined we have monitored 
$\sim 7000$ stars for which variations of $\ga 0.1$\,mag are easily detectable, i.e. we could 
expect up to $\sim 14$ W\,UMa stars among them, which is an upper limit, since our database 
is expected to include a significant fraction of giants. 

\begin{table*}
\centering
\caption{Periodic variables in our monitoring field. Amplitudes measured from the first observing night. \label{t4}}
\begin{tabular}{rcccccrrrl}
\hline
ID & $\alpha$ (J2000) & $\delta$ (J2000) & $K$ (mag) & $H-K$ & $J-H$ & $\Delta$J & $\Delta K$ & P (d) & Comments \\
\hline
1-1141 & 21:26:45.43 & 57:59:42.3 & 13.86 & 0.24 & 0.92 & 0.13 & 0.12 & 0.436 & max/min change \\ 
1-6059 & 21:25:43.09 & 57:50:48:7 & 15.91 & 0.20 & 0.53 & 0.42 & 0.35 & 0.253 & max/min change \\ 
1-6221 & 21:25:21.26 & 57:49:58:8 & 15.16 & 0.22 & 0.73 & 0.34 & 0.35 & 0.129 & max/min change \\ 
\hline
4-1008 & 21:26:57.27 & 58:17:42.5 & 13.69 & 0.16 & 0.52 & 0.08 & 0.07 & 0.433 & max/min change \\ 
4-2728 & 21:25:46.49 & 58:20:36.7 & 15.81 & 0.13 & 0.82 & 0.60 & 0.46 & 0.249 & max/min change  \\ 
4-4066 & 21:26:44.37 & 58:23:19.2 & 11.55 & 0.18 & 0.54 & 0.28 & 0.27 & 0.241 & eclipse?  \\ 
4-5549 & 21:25:35.93 & 58:25:39.2 & 16.89 & 0.30 & 0.52 & 0.38 & 0.48 & 0.213 & max/min change  \\ 
4-5676 & 21:26:23.18 & 58:26:10.9 & 11.58 & 0.28 & 0.49 & 0.28 & 0.25 & 0.311 & \\ 
\hline
\end{tabular}
\end{table*}

The brighter ones could also be pulsating RR\,Lyr stars at distances of $>2$\,kpc. For the fainter objects, this 
would imply unrealistically large distances. Some of the periodic stars could be $\delta$\,Scuti pulsating variables, 
particularly the ones with relatively low amplitudes. The sample may also include spotted, rotating variables 
(BY\,Draconis or FK\,Comae type); 1-1141 and 4-1008 might be candidates for these types of stars. However, the short 
periods, large amplitudes, and min/max changes are atypical for spotted stars. Finally, 4-4066 could alternatively be 
a detached or semi-detached eclipsing binary.

\section*{Acknowledgments}

We would like to thank our referee, Neil J. Evans, for a prompt and helpful revision that helped to improve
the paper. AS would like to acknowledge financial support from the Scottish Universities of Physics Alliance 
SUPA under travel grant APA1-AS110X. 
 
\newcommand\aj{AJ} 
\newcommand\araa{ARA\&A} 
\newcommand\apj{ApJ} 
\newcommand\apjl{ApJ} 
\newcommand\apjs{ApJS} 
\newcommand\aap{A\&A} 
\newcommand\aapr{A\&A~Rev.} 
\newcommand\aaps{A\&AS} 
\newcommand\mnras{MNRAS} 
\newcommand\pasa{PASA} 
\newcommand\pasp{PASP} 
\newcommand\pasj{PASJ} 
\newcommand\solphys{Sol.~Phys.} 
\newcommand\nat{Nature} 
\newcommand\bain{Bulletin of the Astronomical Institutes of the Netherlands}

\bibliographystyle{mn2e}
\bibliography{aleksbib}

\label{lastpage}

\end{document}